\documentclass[12pt]{article}
\usepackage{graphicx}      
\usepackage{setspace}      
\usepackage{enumerate}     
\usepackage{amssymb,amsthm,amsmath,amsfonts,amsbsy}
\usepackage{mathrsfs,syntonly,graphicx,color}
\usepackage{setspace,geometry}
\usepackage{indentfirst}
\usepackage{bbm}
\usepackage{bm}
\usepackage{float}
\usepackage{tabularx}      
\usepackage{tikz}
\usepackage{ctable}        
\usepackage{multirow}      
\usepackage{comment}       
\usepackage[round]{natbib}
\usepackage{caption}
\usepackage[colorlinks, urlcolor=blue, linkcolor=blue,citecolor=blue,hypertexnames=false]{hyperref} 

\usepackage{subcaption} 
\usepackage{soul} 
\usepackage{lineno}
\usepackage{lscape} 
\usepackage{chngcntr} 
\usepackage[labelfont=bf]{caption} 
\usepackage[flushleft]{threeparttable} 
\usepackage{authblk}
\usepackage{diagbox}
\usepackage{dcolumn}
\usepackage{multirow}

\usepackage{multibib}
\newcites{app}{Appendix References}

\title{Political Shocks and Price Discovery in Prediction Markets: Evidence from the 2024 U.S. Presidential Election}
\author{Kwok Ping Tsang and Zichao Yang\thanks{Tsang: Department of Economics, Virginia Tech, Pamplin Hall, Blacksburg, VA 24061; E-mail: byront@vt.edu. Yang: Wenlan School of Business, Zhongnan University of Economics and Law, Wuhan, 430073, China; E-mail: yang\_zichao@outlook.com.}}
\date{\today}

\begin{document}

\maketitle

\begin{abstract}
\noindent What do trading and prices each reveal when political news hits a prediction market? We answer using Polymarket's on-chain ledger around three shocks in the 2024 U.S. presidential election: the Biden--Trump debate, the assassination attempt on Trump, and Biden's withdrawal. Trading rises after every shock, mainly among incumbents with greater prior activity and with larger realized gains on pre-event portfolios. Price adjustment differs across shocks. The debate's price increase largely reverses, the assassination-attempt repricing persists, and Biden's withdrawal generates the heaviest trading with little change in the Trump price. The price response tracks what the news reveals about linked candidates and how much was already anticipated, not the amount of trading. Trading volume measures participation, while belief revision must be read from linked outcome prices and the surprise in the news.
\end{abstract}

~

~


\clearpage
\doublespacing
\section{Introduction}
\label{sec:introduction}

\noindent Prediction-market prices are often read as real-time measures of how breaking news changes beliefs, while trading volume is read as the strength of the market's response. Both readings become difficult in a categorical market. News can bring traders back while reallocating probability across linked outcomes, leaving one candidate's price almost unchanged. The same event then raises two separate economic questions: who supplies the increase in trading, and which candidates' winning probabilities change? We call these the trader response and the price response. This paper uses three political shocks in Polymarket's \href{https://polymarket.com/event/presidential-election-winner-2024}{2024 U.S. presidential election market} to explain why the two responses can diverge.

We study the Biden--Trump debate, the assassination attempt on President Trump, and President Biden's withdrawal from the race. Within four weeks, the same Trump YES market produced reversal, persistence, and almost no price movement despite heavy trading. This contrast provides a within-market comparison that holds the institutional setting fixed while the news and resulting price paths differ. Polymarket's complete ledger on the Polygon blockchain connects those price paths to the trading behind them: it identifies the wallets, order direction, and initiating side of each matched trade. For the trader response, we measure new entry and incumbent activity, then relate the event-time change in each incumbent's trading to prior activity and the realized revaluation of the pre-event portfolio. For the price response, we reconstruct how orders cleared, compare price changes across the linked candidate YES markets, and measure prior anticipation in the Democratic-nominee market.

We find that the trader response comes mainly from the market's existing active traders. New address entry rises after the assassination attempt and Biden's withdrawal, but not after the debate. Incumbent activity rises after all three shocks. Greater prior activity is associated with a larger post-shock increase at every event, most strongly at the assassination attempt and withdrawal. At those two events, realized gains on pre-event portfolios are also associated with a larger increase in trading. Active incumbents thus supply the trading response to every shock, while the extent of new entry and the role of realized gains vary with the event.

The price response follows a different pattern. The debate's initial increase largely reverses, the assassination attempt's increase persists, and Biden's withdrawal generates the heaviest activity but little price movement. The withdrawal is the sharpest case, and it raises two questions: how heavy flow cleared near an unchanged price, and why traders were willing to meet there. The order-level ledger answers the first: a concentrated group of traders cut Trump exposure against dispersed resting demand, and most Trump YES shares supplied to buyers were newly minted. The linked candidate markets answer the second: the news reallocated probability from Biden YES to Harris YES rather than repricing the Trump-versus-Democrat contest. By the announcement, the \href{https://polymarket.com/event/democratic-nominee-2024}{Democratic-nominee market} had assigned a 0.70 probability to a Biden replacement, so much of that substitution was already priced. The comparison events sharpen the contrast. The assassination attempt moved Biden YES and Harris YES down together as Trump YES rose, the signature of news about the Trump-versus-Democrat contest, and a rolling price-impact decomposition tracks a stronger permanent component behind its persistence and a stronger transitory component behind the debate's reversal. Across the three shocks, the Trump price follows what the news reveals about linked outcomes and how much was already anticipated, not the amount of trading alone.

We make two contributions to the study of information aggregation in prediction markets. First, we show who supplies the trader response to public news. Wallet-level evidence reveals that shocks draw different degrees of new entry, while the immediate incumbent response is concentrated among traders already active before the event and, at the assassination attempt and withdrawal, among traders with realized portfolio gains. This result connects aggregate event volume to the participants whose activity generates it. Second, we explain why the price response can diverge from heavy trading in a categorical market. At Biden's withdrawal, the ledger and linked markets reveal two parts of the same episode: flow cleared through dispersed resting demand and newly minted shares, while the news mainly reallocated probability within the Democratic field and was largely anticipated. Together, these results change how prediction-market activity should be read. Volume is evidence of the trader response, but belief revision must be inferred from linked outcome prices and the surprise in the news.

\subsection{Related literature}
\label{sec:related}

Prediction markets are studied as information-aggregation devices, with work on forecast accuracy, market design, and manipulation \citep{wolfers2004prediction, arrow2008promise, SnowbergWolfersZitzewitz2013, RhodeStrumpf2004, berg2008accuracy, berg2008results, RhodeStrumpf2007Manipulation, BergRietz2014}. Recent studies examine the 2024 election setting directly. \citet{ChernovElenevSong2025} estimate state-level voter preferences from polls, fundamentals, and prediction-market prices. \citet{SethiEtAl2025} compare prediction-market prices with statistical forecasts, and \citet{NgEtAl2025} study price discovery across trading platforms. \citet{tsang2026anatomy} document the design, volume composition, and market-wide evolution of Polymarket's presidential market. We study how one categorical market processes timed public shocks, connecting the wallets that trade afterward to adjustment across linked candidate YES prices.

On the trader side, our analysis relates to work on heterogeneity in trading intensity, learning, and attention \citep{barberodean2000trading, seru2010learning, barber2022attention}. \citet{RothschildSethi2016} document heterogeneous strategies in an earlier presidential prediction market. We add a measure of each trader's exposure to the news itself. Revaluing pre-event portfolios at a common realized event-price vector makes exposure comparable across the three shocks.

On the price side, our analysis draws on the market-microstructure literature on signed order flow and price impact \citep{kyle1985continuous, GlostenMilgrom1985, hasbrouck1991measuring, hasbrouck1995one, easley1997information, glostenharris1988}. This literature commonly infers the trade initiator with classifiers in the spirit of \citet{lee1991inferring}. Polymarket's on-chain ledger instead identifies order direction and the initiating side directly, so the standard classifier can be audited against the ledger. The ledger also reveals whether shares came from existing inventory or were minted as a new YES--NO pair. Minting parallels share creation in other financial markets \citep{brown2021}, but it enforces the YES--NO parity rather than anchoring the market to a fundamental value.

Finally, two explanations can account for heavy trading with a small price change. Different interpretations of public news can generate volume without much price movement \citep{kandelpearson1995}. Prices should also respond only to the unanticipated component of an announcement \citep{fama1969adjustment, kuttner2001monetary}. The complete ledger identifies the asymmetric form of disagreement at the withdrawal, while the Democratic-nominee market provides a traded measure of anticipation. Together, the ledger and nominee price distinguish disagreement about the news from anticipation of the announcement.

The rest of the paper is organized as follows. \autoref{sec:data} describes the market, data, and measurement. \autoref{sec:react} studies entry and incumbent activity. \autoref{sec:heterogeneity} identifies which incumbent traders supply the immediate response. \autoref{sec:price_discovery} studies clearing, linked price adjustment, anticipation, and the persistence of the price response. \autoref{sec:conclusion} concludes.

\section{Institutional Background, Data, and Measurement}
\label{sec:data}

\noindent The Polymarket setting pairs precisely timed political shocks with two records that are seldom available together: prices across linked candidate markets and the wallets behind every executed order. This section describes the market that generates these records, the dataset we construct from them, and the event-time measures used throughout the paper.

\subsection{Polymarket market structure}

Polymarket is a hybrid-decentralized prediction market that combines off-chain matching with on-chain settlement. One kind of prediction market operating on Polymarket is the categorical market: several outcomes are mutually exclusive and collectively exhaust the event, and each outcome is traded through YES and NO shares. Orders are submitted and matched off chain in a centralized limit order book, while completed trades settle on Polygon through smart contracts using USDC, a stablecoin pegged one-for-one to the U.S. dollar, as collateral. This design gives traders a fast and low-friction trading interface, but it also means that our data reveal executed trades rather than the full standing order book.

The main form of trading is peer-to-peer exchange of existing shares. A trader submits an order, Polymarket matches compatible orders in the central limit order book, and matched orders then go on chain to the relevant exchange contract for atomic execution. Completed matches are therefore observable in blockchain logs even though order submission and matching are not. For the presidential market, this architecture lets us recover transaction prices, quantities, timestamps, and counterparties at high frequency from the on-chain ledger.

Besides peer-to-peer trading, Polymarket's protocol also supports position split, merge, and conversion functions. These functions create or burn bundles of YES and NO shares, or convert NO shares into economically equivalent YES shares in complementary outcomes, and they are central to arbitrage and capital efficiency in categorical markets. Splitting, in particular, lets the exchange fill a buyer of YES and a buyer of NO against each other by minting a new share pair against their combined collateral, a mechanism that plays a central role in \autoref{sec:clearing}. Merging is the mirror image: the exchange can fill a seller of YES against a seller of NO by burning an existing pair back into its collateral. These functions enforce pricing consistency both within a given candidate market and across mutually exclusive outcomes. A more detailed discussion of Polymarket's market design is provided in \citet{tsang2026anatomy}.

The 2024 U.S. Presidential Election Winner market is a categorical market with 17 mutually exclusive outcomes. We focus on Trump, Biden, and Harris because those three candidate markets account for nearly all economically relevant trading activity in our sample.

\subsection{Data and variable construction}

\subsubsection{Data collection}

Our dataset is built from the matched-trade logs emitted by Polymarket's exchange contracts. For every executed match, we observe the traded value in USDC, token quantity, timestamp, and the wallet addresses on both sides of the trade. Together, these logs give us a wallet-resolved transaction history that connects event-time price adjustment to the traders behind it. The empirical analysis focuses on the \href{https://polymarket.com/event/presidential-election-winner-2024}{Presidential Election Winner 2024 market}, whose matched trade logs are emitted by the \emph{NegRisk CTFExchange} contract. We collect the relevant on-chain logs directly from the Polygon blockchain via the \href{https://polygon-rpc.com/}{Polygon RPC} endpoint.

The sample runs from January 5, 2024, when the market launched, through November 6, 2024, just before Fox News first projected the election winner. Over this period, we collect 3,654,974 matched trades involving 232,469 unique addresses in the Trump, Biden, and Harris markets. For the anticipation analysis in \autoref{sec:anticipation} we additionally collect the exchange logs of the \href{https://polymarket.com/event/democratic-nominee-2024}{Democratic-nominee market}, in particular the Biden-nominee and Harris-nominee markets, over the same window.

\subsubsection{Filtering platform and advanced-operator addresses}

Some wallet addresses in the raw data reflect platform mechanics, arbitrage, or inventory management rather than ordinary speculative trading about the final outcome. We identify these addresses and treat them separately using the following heuristic filters.
\begin{enumerate}
    \item \emph{Known Polymarket addresses.} If an address is known to be controlled by Polymarket, we label it as such.
    \item \emph{Position conversion activity.} Addresses that conduct \emph{Position Conversion} via the smart contract \emph{NegRisk Adapter} are flagged as advanced operators, as this activity is commonly associated with arbitrage or inventory management by advanced users.
    \item \emph{Negative token holdings.} Using our reconstructed token holding dataset from exchange-trade logs, we flag addresses whose token holdings have ever turned negative for any candidate, which typically indicates off-exchange inventory changes or conversion activity not captured in the trade logs emitted by \emph{NegRisk CTFExchange}.
    \item \emph{Behavioral market making.} Addresses that trade in large volume on both sides of the order book and end each trading day with little net position change are flagged as behavioral market makers. This criterion captures pure order-book intermediaries that never interact with the \emph{NegRisk Adapter} and would therefore escape the conversion-based filters.
\end{enumerate}
Negative token holdings arise for two reasons. First, position conversions executed via the \emph{NegRisk Adapter} do not appear in the \emph{NegRisk CTFExchange} trade logs used to reconstruct our token holdings dataset. The subsequent sales of converted tokens therefore appear as sales of inventory ``never purchased'' in our dataset. Second, addresses may transfer tokens across wallets, for example from a cold wallet to a hot wallet. These transfers occur outside the exchange and are not observable in our trading dataset. Subsequent exchange trades can therefore create negative reconstructed holdings even when the address previously received the tokens elsewhere. In our data, only seven addresses show transfer-driven negative holdings without also exhibiting conversion activity. We exclude all seven from the analysis.

\begin{figure}[hbtp!]
	\centering
	\includegraphics[width=13cm]{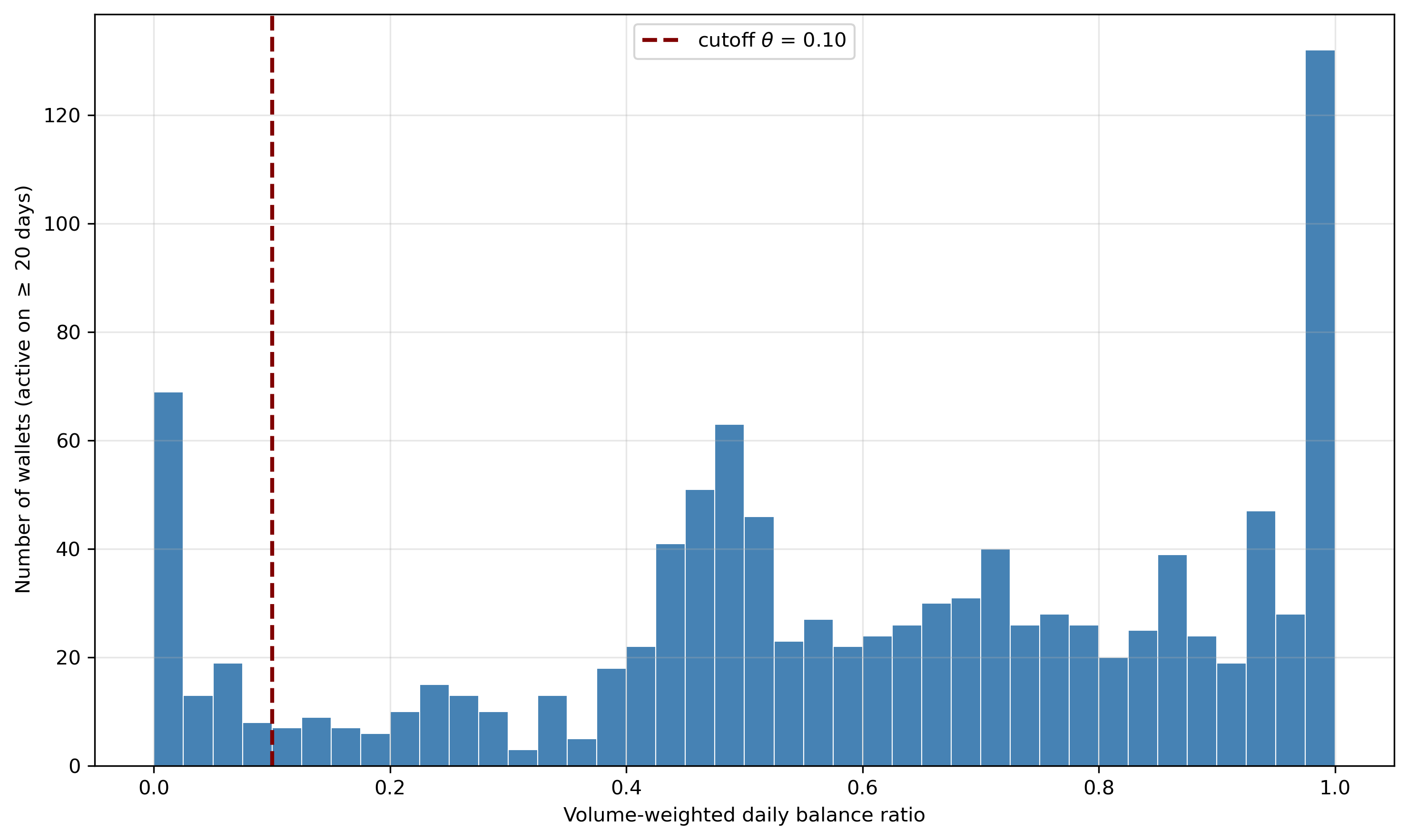}
	\caption{Distribution of volume-weighted daily balance ratios}
	\label{fig_mm_balance_distribution}
	\vspace{0.25em}
	\begin{minipage}{0.98\linewidth}
	\footnotesize \raggedright \textit{Notes:} The sample is the 1,085 addresses active on at least 20 days in the Trump, Biden, and Harris YES and NO markets. For each address, the ratio is the sum over days of absolute net changes in exposure-direction tokens divided by gross tokens traded. The dashed line marks the 0.10 cutoff for behavioral market makers.
	\end{minipage}
\end{figure}

For the behavioral market-making filter, each address $i$ receives a volume-weighted daily balance ratio
\begin{equation}
\text{DailyBalance}_i=\frac{\sum_{d}\sum_{c}\left|\text{net tokens}_{i,d,c}\right|}{\sum_{d}\sum_{c}\text{gross tokens}_{i,d,c}},
\end{equation}
where net tokens$_{i,d,c}$ is the address's net change in exposure to candidate $c$ winning on day $d$, aggregated across that candidate's YES and NO markets, with purchases in the YES market and sales in the NO market counted as positive, and gross tokens$_{i,d,c}$ is the total token quantity the address traded across that candidate's YES and NO markets on that day. We aggregate the two markets of the same candidate because purchases of YES and NO offset into a riskless position, so an intermediary holding both legs carries no net exposure. A market maker that ends each day close to flat has a ratio near zero. A directional bettor has a ratio near one. A trader who accumulates a position in one week and exits in a later week is one-sided within each day and is therefore not flagged. We flag addresses that are active on at least 20 days and have a daily balance ratio of at most 0.10. The cross-sectional distribution of the ratio, shown in \autoref{fig_mm_balance_distribution}, supports this cutoff. Among persistently active addresses, the distribution is sharply bimodal, with one mode at zero and the other near one, and the region between roughly 0.05 and 0.25 is sparsely populated, so the flagged set is insensitive to the exact cutoff within that range. The criterion follows the literature that identifies non-designated market makers from their trading behavior, namely large volume with inventory held close to zero \citep{menkveld2013, kirilenko2017}.

We also remove clusters of addresses that transfer tokens among themselves off exchange. These addresses appear to be controlled by the same underlying entity for operational or security reasons. Removing these clusters reduces double-counting of economically distinct traders and contamination from operational wallet management. After applying the first three filters, we classify 2,176 addresses as advanced operators and identify 355 additional addresses that are likely linked to the same underlying entities. The behavioral market-making filter flags 109 addresses, of which 91 are not captured by any of the other filters.

Accordingly, the trader-level analyses in \autoref{sec:react} and \autoref{sec:heterogeneity} exclude the flagged addresses and clusters, so that entry and event-time activity describe ordinary traders. By contrast, the market-level flow and clearing analyses in \autoref{sec:price_discovery} include all wallets because the flagged intermediaries are part of the clearing mechanism.

\subsection{Order-level trade direction from the on-chain ledger}
\label{sec:flow_measure}

The exchange logs in Polymarket's on-chain ledger contain only executed trades: we observe who traded, when, and at what price, but not the standing order book. The absence of standing orders leads us to transaction-based measures of activity, order flow, and price impact rather than quoted spreads or displayed depth. The unit of observation is an order fill, which appears in the ledger as an \texttt{OrderFilled} event. Each fill entry gives the wallet address for one filled order and the two asset legs of that fill. When $\texttt{makerAssetId}=0$, the wallet pays USDC and buys the token in \texttt{takerAssetId}. When $\texttt{takerAssetId}=0$, the wallet delivers the token in \texttt{makerAssetId} and sells it. Buying YES or selling NO raises exposure to the candidate winning. Selling YES or buying NO lowers it. All wallet-level measures aggregate these order fills by wallet, token, and time.

To construct order flow as initiator buys minus initiator sells, we also identify which order crossed the book. The ledger distinguishes the incoming order from the resting orders within each match. \autoref{appendix_trade_direction} illustrates this classification with representative matches and validates it across the ledger. As a robustness check, we re-estimate every flow-based result under the standard tick rule of \citet{lee1991inferring}, which infers the initiator from price movements. The two methods agree on 97\% of transactions, and every event-window net-flow sign is unchanged (\autoref{tab_dual_signing}).

\subsection{Event design and measurement}
\label{sec:measurement}

Our analysis centers on three political events with observable time anchors: the start of the Biden--Trump debate (2024-06-28, 1:00 UTC), the assassination attempt on President Trump (2024-07-13, 22:11 UTC), and President Biden's withdrawal from the race (2024-07-21, 17:46 UTC). These anchors let us aggregate raw trades to 5-minute bins and use the same event-time definition across sections.\footnote{We experimented with alternative bin widths of 1, 5, 10, 30, and 60 minutes. The 5-minute bin is short enough to capture the immediate response to new information while still smoothing the noise created when one order fills as several trades.}

The event design uses two horizons for distinct questions. For the immediate trader response, we use thirty minutes before to thirty minutes after each shock.\footnote{Elon Musk endorsed Trump at 22:45 UTC, and Biden endorsed Harris at 18:13 UTC. The thirty-minute window excludes most of the contamination from these subsequent events while retaining enough observations for the analysis.} The panel relates the change in each trader's 5-minute activity to prior activity and realized portfolio exposure. Event revaluation applies the candidate YES price changes over the first thirty post-event minutes to portfolios fixed before the event window, so it measures each trader's realized exposure to the common shock. The debate released information over ninety minutes, so its thirty-minute change captures the early response rather than the full debate. By contrast, we use a three-hour horizon to describe whether the initial price response persists or reverses. \autoref{tab_measurement_map} summarizes the construction.

\section{Event-Driven Trading Responses}
\label{sec:react}

\noindent We begin by showing when trading activity and winning probabilities shift around the three shocks. We then separate the immediate trading increase into new address entry and abnormal trading by incumbents. The two margins need not move together, so distinguishing them shows whether an event reaches beyond the existing trader base or mainly brings back traders already present.

\subsection{Market-level price and market activity dynamics}

\begin{figure}[hbtp!]
\centering
\begin{subfigure}{\linewidth}
	\centering
	\includegraphics[width=10.5cm]{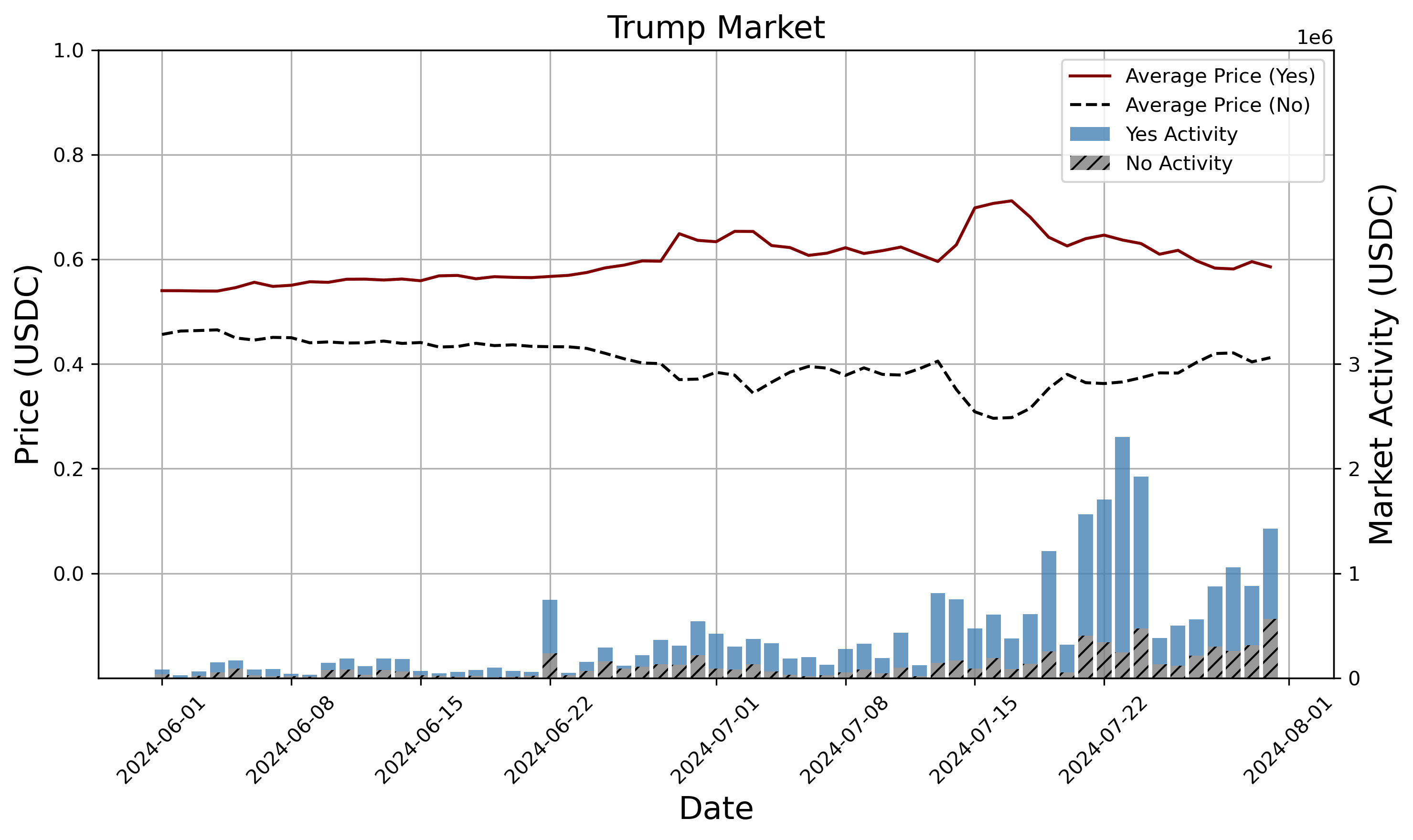}
\end{subfigure}
\begin{subfigure}{\linewidth}
	\centering
	\includegraphics[width=10.5cm]{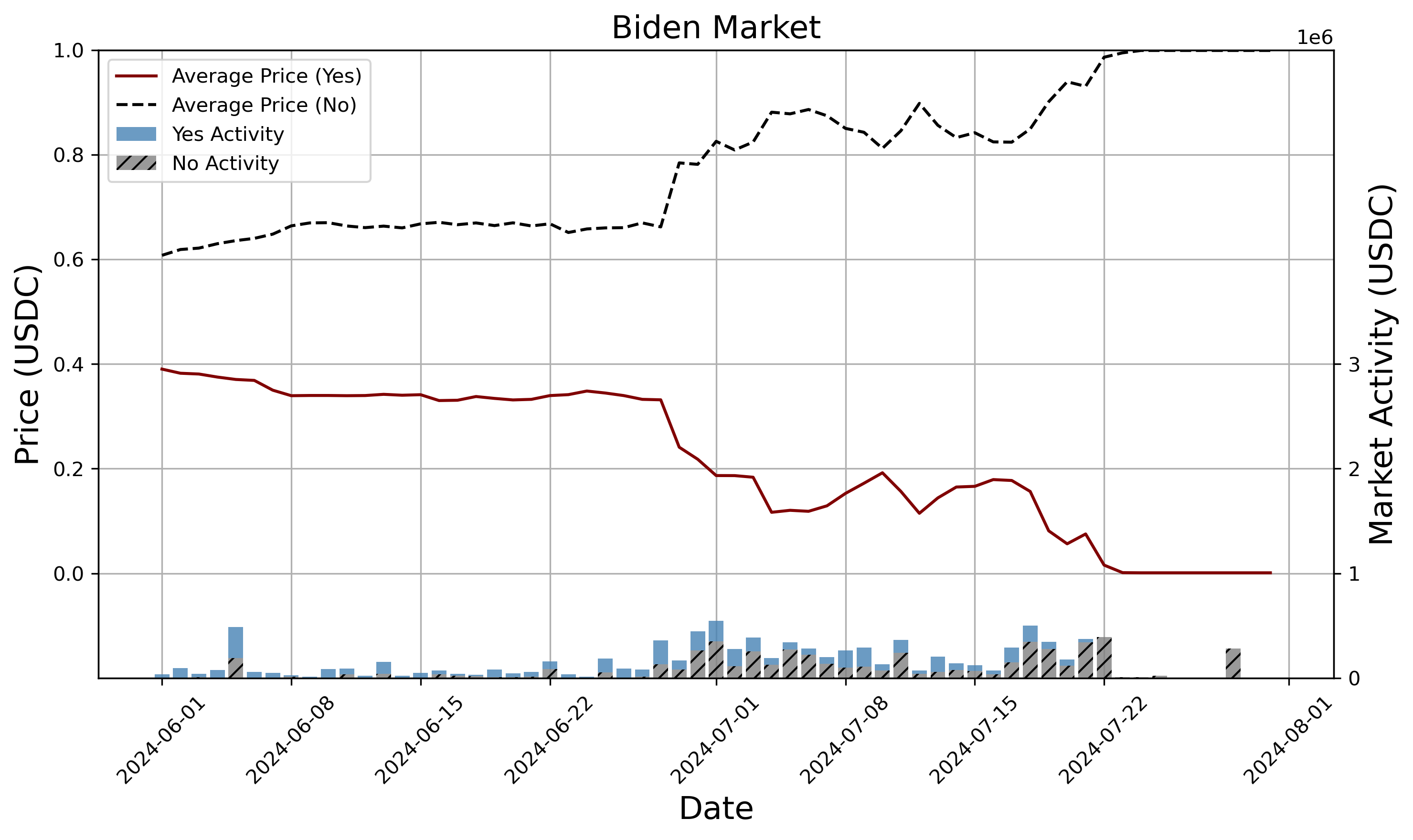}
\end{subfigure}
\begin{subfigure}{\linewidth}
	\centering
	\includegraphics[width=10.5cm]{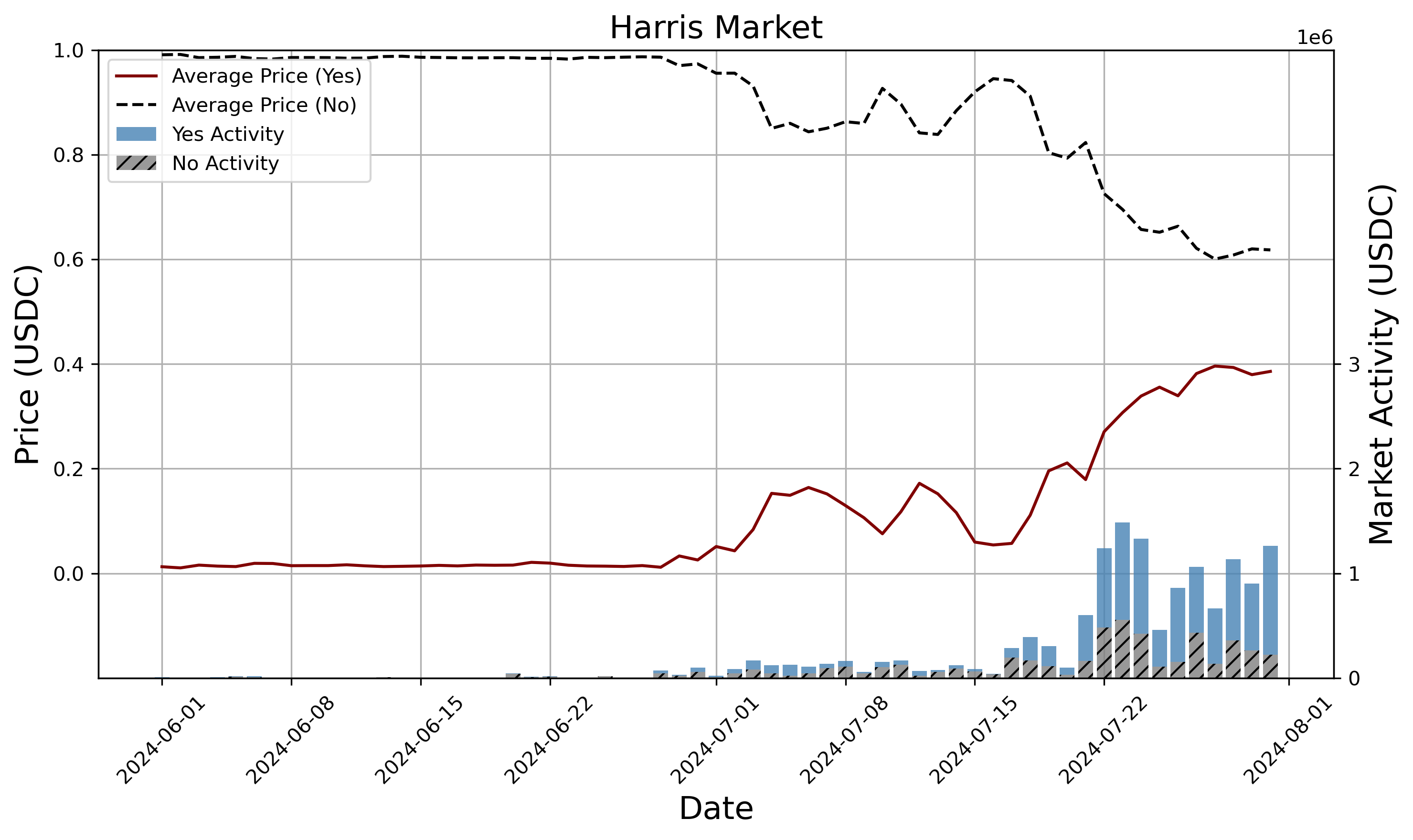}
\end{subfigure}
\caption{Prices and activity across candidate markets}
	\label{fig_price_volume}
\vspace{0.25em}
\begin{minipage}{0.98\linewidth}
\footnotesize \raggedright \textit{Notes:} Lines show daily average YES and NO prices. Bars show daily market activity, defined as trading volume plus the larger of minting and burning volume \citep{tsang2026anatomy}.
\end{minipage}
\end{figure}

\autoref{fig_price_volume} plots the evolution of prices and market activity in the Trump, Biden, and Harris prediction markets over the period spanning the three events. Polymarket's on-chain ledger records position splits and merges alongside peer-to-peer trades, so raw ledger volume mixes trading with mechanical share operations. We therefore measure activity with the decomposition proposed in \citet{tsang2026anatomy}. With this measure, the figure establishes two facts used below: probability was already shifting across candidate markets before Biden's withdrawal, and the Trump market remained the main venue for trading through most of the sample.

Biden YES falls after the June 28 debate, indicating that traders read the performance as materially damaging Biden's prospects. Harris YES begins rising well before July 21, consistent with traders assigning increasing probability to candidate replacement before Biden formally withdrew. The joint movement in Trump YES and Harris YES is consistent with a growing probability of a Trump--Harris contest. These pre-withdrawal price shifts preview the anticipation results of \autoref{sec:anticipation} and help explain why the withdrawal reallocates probability from Biden YES to Harris YES while moving the Trump YES price only slightly.

Trading activity follows the same shift across candidate markets. The Trump market remains the main venue before July 21 and therefore provides most of the transaction-level evidence below. After Biden's withdrawal, activity shifts toward Harris.

\subsection{Extensive margin: new trader entry}

\begin{figure}[hbtp!]
	\centering
	\includegraphics[width=15cm]{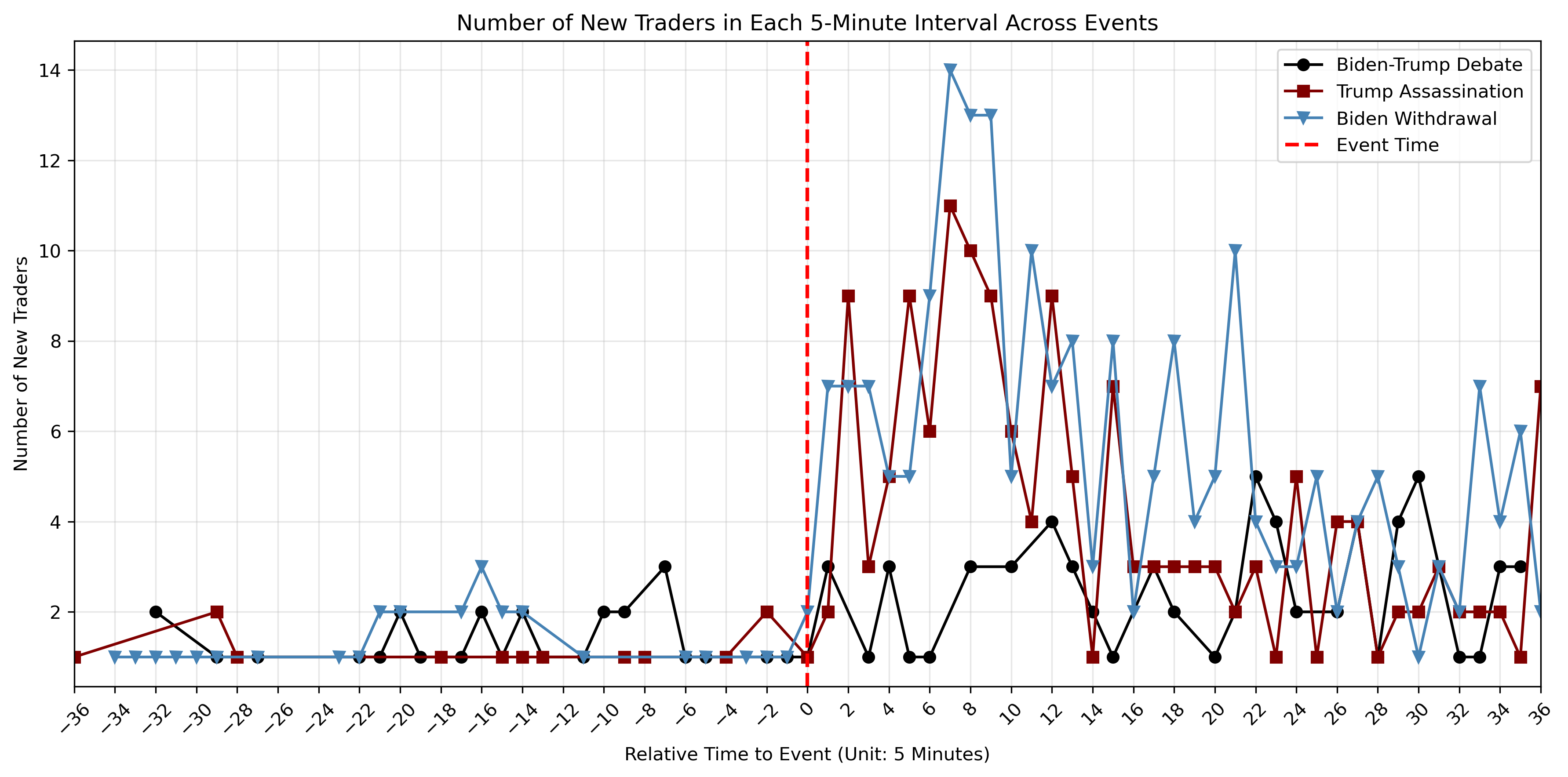}
	\caption{New trader entry around political shocks}
	\label{fig_newcomers}
	\vspace{0.25em}
	\begin{minipage}{0.98\linewidth}
	\footnotesize \raggedright \textit{Notes:} New traders are wallet addresses whose first observed trade occurs in the indicated 5-minute event-time bin. New addresses need not be new people. See the sybil caveat in the text.
	\end{minipage}
\end{figure}

New entry differs across events (\autoref{fig_newcomers}). The assassination attempt and Biden's withdrawal generate increases in first-time trading, consistent with these events drawing outside attention to the market. The debate generates no comparable entry response, perhaps because Polymarket was still unfamiliar to much of the public in late June. New addresses are not necessarily new people because creating wallets is inexpensive, so the counts are an upper bound on genuine entry. The contrast shows why entry and incumbent activity must be separated: the debate drew few newcomers but still brought existing traders back to the market in response to Biden's performance.

\subsection{Intensive margin: incumbent trading response}

For incumbents, the question is whether the shock moves each trader's activity away from its own pre-event level. We therefore implement a standard event-study design at 5-minute frequency \citep{brown1985using, mackinlay1997event}. For each event $e$, we re-index trading into event-time bins around $t=0$ and measure trader-level volume and frequency. Trading participation is an indicator equal to 1 if the trader is active in a given 5-minute bin, because a single limit order can be split into multiple trades. We define abnormal activity relative to a trader-specific baseline from the three hours before the event. We transform volume and frequency with the inverse hyperbolic sine, retaining zero-trading bins while reducing the influence of large observations. \autoref{appendix_event_study} gives the exact definitions.

\begin{figure}[hbtp!]
	\centering
	\begin{subfigure}{1\linewidth}
		\centering
	\includegraphics[width=1\textwidth]{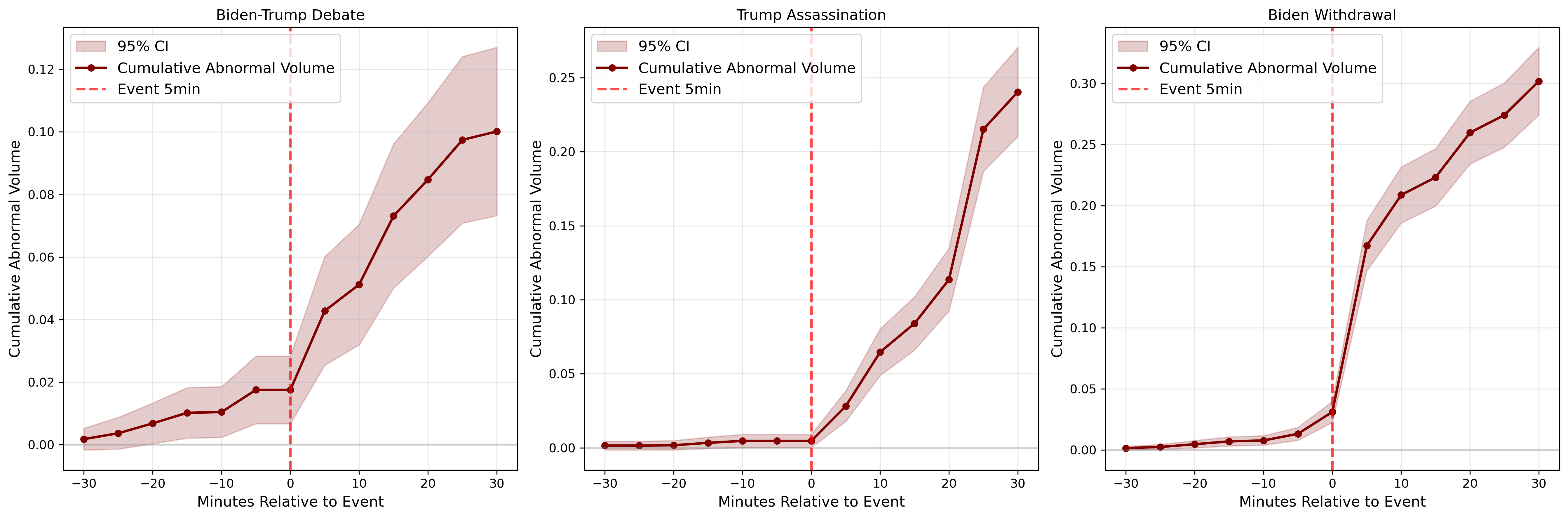}
\caption{Trading volume}
\end{subfigure}
\begin{subfigure}{1\linewidth}
	\centering
	\includegraphics[width=1\textwidth]{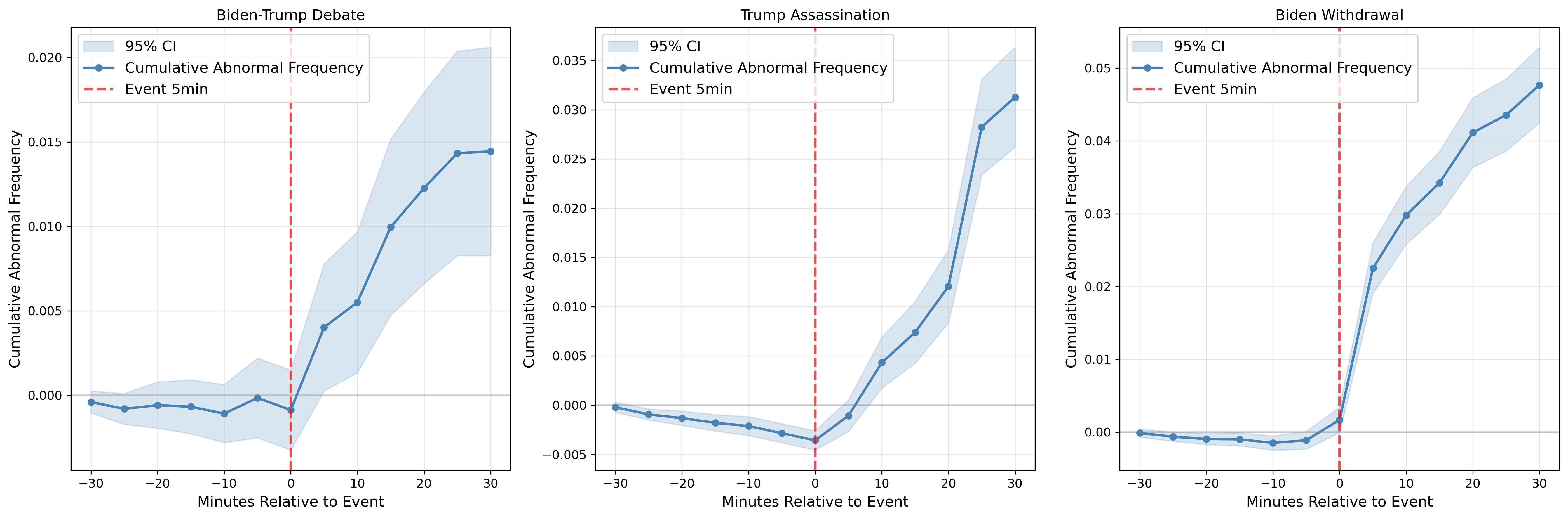}
\caption{Trading frequency}
\end{subfigure}
\begin{subfigure}{1\linewidth}
	\centering
	\includegraphics[width=1\textwidth]{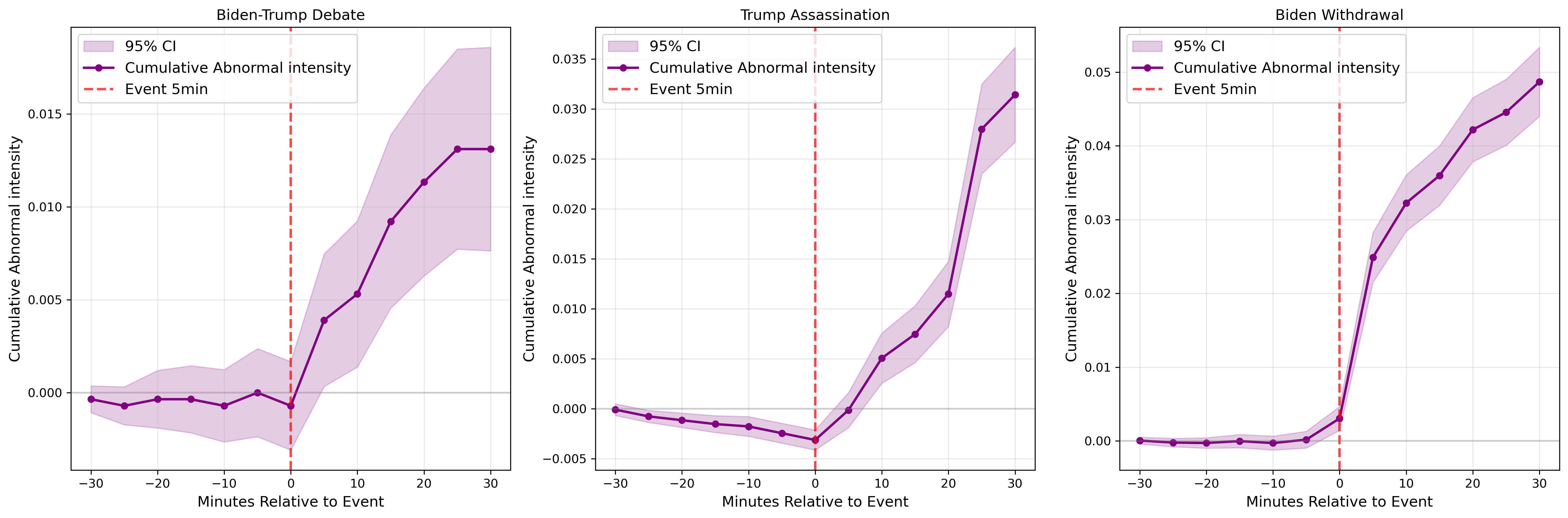}
		\caption{Trading participation}
\end{subfigure}
\caption{Cumulative abnormal incumbent trading activity}
	\label{fig_intensive}
	\vspace{0.25em}
	\begin{minipage}{0.98\linewidth}
	\footnotesize \raggedright \textit{Notes:} Panels show cumulative abnormal incumbent trading activity in 5-minute event time from -30 to +30 minutes, with 95\% confidence bands. The vertical line marks event time. Abnormal activity is measured relative to each trader's pre-event baseline (see \autoref{appendix_event_study}).
	\end{minipage}
\end{figure}

Incumbent trading rises after all three shocks (\autoref{fig_intensive}). The assassination attempt and Biden's withdrawal produce the largest responses, while the debate produces a smaller increase. The increase persists beyond the first bin, so incumbent trading is not confined to the first moments after the news. Placebo tests matched on weekday and clock time show that comparable increases in volume, frequency, and participation are rare outside the event days (\autoref{fig_placebo_intensive_margin}).

New entry rises after the assassination attempt and withdrawal, but all three events bring incumbent traders back to the market. We next ask which incumbents supply that increase.

\section{Which Traders Supply the Response?}
\label{sec:heterogeneity}

\noindent Incumbent activity rises after all three shocks, but the aggregate series cannot show which wallets came back. We use the wallet-level panel to ask whether the increase is larger among traders who were more active beforehand and among those whose pre-event portfolios were revalued by the shock. We also control for single-market participation and contrarian or momentum type.

\subsection{Prior activity and realized exposure}

The analysis uses 5-minute event time from thirty minutes before through thirty minutes after each shock. \emph{Trad Vol High} equals one when the trader's volume during the preceding month is above the median among traders active in that month. Because prior-month frequency has a mass point at one trade, \emph{Trad Freq High} equals one when the trader traded more than once during that month. Both characteristics are therefore predetermined at the shock.

In addition to prior activity, realized portfolio exposure captures how the same event affects traders holding different candidate positions. Let $h_{iT}$, $h_{iB}$, and $h_{iH}$ denote trader $i$'s net positions, YES tokens minus NO tokens, in the Trump, Biden, and Harris markets, fixed before the event window. We revalue these positions at the common candidate YES price changes over the first thirty post-event minutes:
\begin{equation}
MtM_{ie}= h_{iT}\,\Delta p^{e}_{T} + h_{iB}\,\Delta p^{e}_{B} + h_{iH}\,\Delta p^{e}_{H}.
\end{equation}
Here $\Delta p^{e}_{T}$, $\Delta p^{e}_{B}$, and $\Delta p^{e}_{H}$ are the changes in the Trump, Biden, and Harris YES prices over the first thirty minutes of event $e$. The resulting $MtM_{ie}$ is the dollar paper gain or loss on trader $i$'s fixed portfolio under the event's realized price vector. \emph{MtM Loss} equals one when the revaluation is below $-1$ USDC, and \emph{MtM Gain} equals one when it is above $1$ USDC. \autoref{table_trader_characteristics} defines the full set of variables, and \autoref{table_summary_trader_characteristics} reports their sample shares. We use these realized-exposure indicators to sort traders, so the coefficients describe correlations between portfolio exposure and event-time activity.

The variables enter stacked panel regressions with trader-by-event and relative-event-time fixed effects:
\begin{equation}
y_{iet}=\alpha_{ie}+\gamma_t+\bm{\beta}'\bm{X}_{ie}\times \text{Post Event}_t+\epsilon_{iet},
\end{equation}
where $y_{iet}$ is the inverse hyperbolic sine of trader $i$'s trading volume or frequency in event $e$ and 5-minute bin $t$. The same trader therefore contributes a different panel unit in different events. The vector $\bm{X}_{ie}$ contains the characteristics in \autoref{table_trader_characteristics}, and $\text{Post Event}_t$ equals one in bins at and after the event. The trader-by-event effects $\alpha_{ie}$ absorb each trader-event's average activity, and the relative-time effects $\gamma_t$ absorb the common event-time profile. Standard errors are clustered by trader-event. \autoref{tab_panel_volume_5min} and \autoref{tab_panel_freq_5min} report the volume and frequency estimates.

\subsection{The active traders supply the cross-event response}

We find that traders who were more active before the shock increase their activity more afterward (\autoref{tab_panel_volume_5min}, \autoref{tab_panel_freq_5min}). In the full specifications, \emph{Trad Vol High} and \emph{Trad Freq High} are positive and significant for both volume and frequency. One concern is that a single order filling as several trades inflates measured volume and frequency. The participation specification in \autoref{tab_panel_intensity_5min}, which records only whether a trader is active in a bin, is immune to this inflation and gives the same result.

The pooled estimates could in principle be driven by one dominant event. The event-specific regressions in \autoref{appendix_event_specific_panel} show instead that the same pattern appears at each shock: prior volume's coefficient is positive in every event-specific volume, frequency, and participation regression, with the largest estimates at the assassination attempt and withdrawal. Prior frequency is also positive in all three event-specific frequency regressions. The market's existing active traders therefore supply much of the immediate response, while the smaller debate coefficients indicate a weaker separation between recently active and other incumbents.

Furthermore, traders with realized gains become more active after the news in the pooled sample. \emph{MtM Gain} is positive and significant in the full volume, frequency, and participation specifications, while \emph{MtM Loss} is close to zero. The event-specific estimates show where this pooled pattern comes from. Gain-side exposure is positive at the assassination attempt and withdrawal, while the debate response is concentrated on the loss side. Realized exposure therefore helps describe which side becomes more active in each event, while prior activity supplies the more stable cross-event pattern.

The trader results establish who supplies the immediate response. The next section asks how the withdrawal's heavy flow cleared near an unchanged Trump YES price and why that price moved so little.

\section{Price Adjustment After Political Shocks}
\label{sec:price_discovery}

\begin{figure}[hbtp!]
	\centering
	\includegraphics[width=\textwidth]{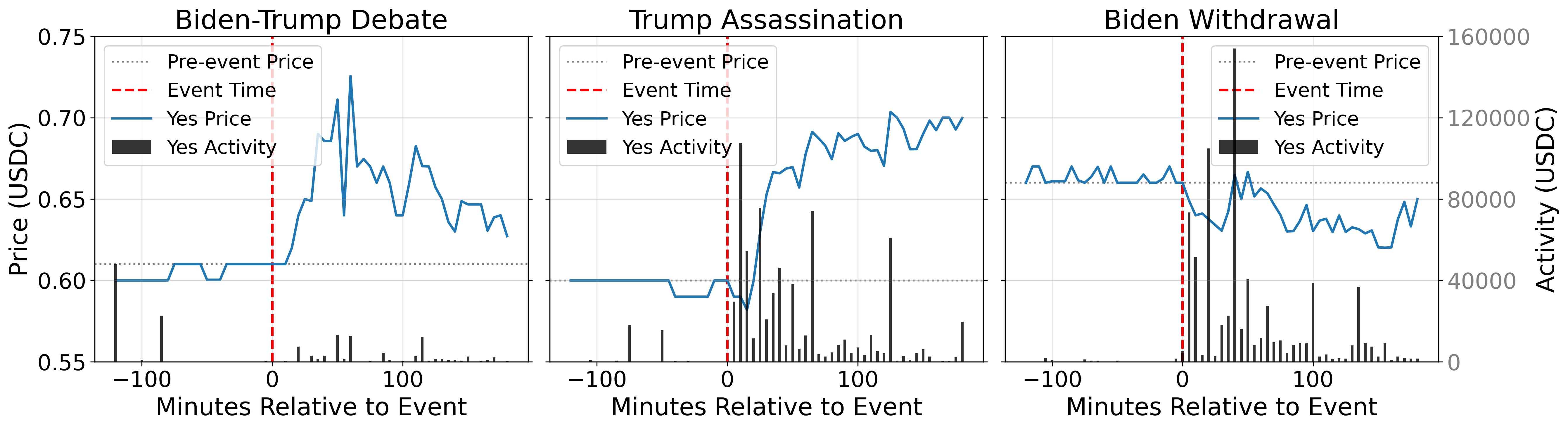}
	\caption{Trump YES adjustment paths around political shocks}
	\label{fig_trump_tokens_price_tx_vol_events}
	\vspace{0.25em}
	\begin{minipage}{0.98\linewidth}
	\footnotesize \raggedright \textit{Notes:} Panels plot the 5-minute Trump YES price from two hours before to three hours after each event. The dotted horizontal line is the pre-event price, the dashed vertical line marks event time, and gray bars show Trump YES activity.
	\end{minipage}
\end{figure}

\noindent How does trading activity affect the Trump YES price? A standard regularity in asset markets links trading volume to the size of price changes, because both rise with the rate of information arrival \citep{karpoff1987}. Biden's withdrawal breaks that link. \autoref{fig_trump_tokens_price_tx_vol_events} plots the Trump YES price and Trump YES activity in the $[-2,+3]$ hour window around each shock and shows three different adjustment paths. The debate raises the price sharply, then gives back most of its peak increase within three hours. The assassination attempt raises it, and the move persists and grows. The withdrawal leaves the price almost exactly where it began. \autoref{fig_activity_decomp} shows that the withdrawal nevertheless generates the most total YES-market activity across the linked candidate markets in the first thirty minutes. Reversal, persistence, and a flat price in the same market are the patterns this section explains, and the price paths alone do not reveal why they differ.

Two questions organize the withdrawal analysis. The first is how heavy Trump flow cleared near an unchanged price. The order-level ledger identifies which traders demanded immediacy, who supplied the resting side, and whether the shares came from existing inventory or new minting. The second is why traders were willing to meet near that price. The linked candidate YES prices show what repriced, and the Democratic-nominee market shows how much of the news was already anticipated. After answering these two questions, we return to the comparison events and ask why the debate's initial move reversed while the assassination response persisted.

\subsection{How the withdrawal flow cleared}
\label{sec:clearing}

Heavy trading reached the Trump YES market at the withdrawal, yet it cleared at an almost unchanged Trump YES price. The order-level ledger shows that taker pressure did not come from similarly sized groups on the two sides. In the first thirty minutes, 94\% of taker-order flow reduced Trump exposure (\autoref{tab_clearing_anatomy}). Thirteen wallets initiated \$160,600 of Trump YES sales and sixteen initiated \$102,100 of Trump NO purchases. Only \$15,500 of taker orders raised Trump exposure, mostly through Trump YES purchases spread across twenty-seven wallets. Tick-rule signing gives the same direction of net pressure (\autoref{tab_dual_signing}).

The two sides of this clearing are populated very differently, a contrast a price-only study cannot see. Classifying every wallet by its net exposure change gives 261 wallets raising Trump exposure and 37 reducing it. The five largest reducing wallets account for 87\% of the reducing flow. The reducers execute 89\% of their volume as taker orders, while the raisers execute 86\% from the resting side. A concentrated group of large wallets therefore demanded an exit from Trump exposure, while a broad group raised exposure mainly through resting orders.

\begin{figure}[hbtp!]
	\centering
	\includegraphics[width=\textwidth]{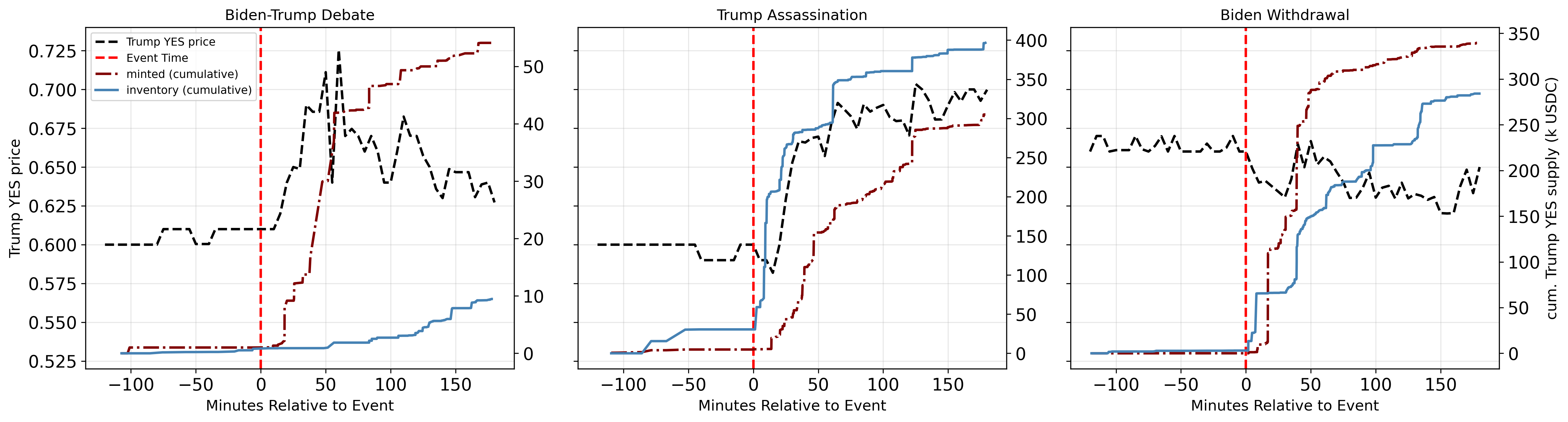}
	\caption{Trump YES supply and price around political shocks}
	\label{fig_supply_source}
	\vspace{0.25em}
	\begin{minipage}{0.98\linewidth}
	\footnotesize \raggedright \textit{Notes:} The dashed line is the Trump YES price on the left axis. The right axis splits cumulative Trump YES supplied to buyers into newly minted shares (dash-dotted) and shares sold from existing inventory (solid).
	\end{minipage}
\end{figure}

Moreover, minting supplied the shares needed for this configuration to clear. When a taker buys Trump NO and the standing book contains bids for Trump YES, the exchange can fill both sides by minting a new YES--NO pair against their combined collateral. Using exact thirty-minute windows, 66\% of the Trump YES supplied to buyers was newly minted at the withdrawal, compared with 14\% at the assassination attempt, where the price rose and demand was filled mainly from inventory (\autoref{tab_fill_mechanism}). \autoref{fig_supply_source} shows the corresponding 5-minute event-time paths. Because a pair can be minted at par whenever the two sides' prices sum to one, no dealer inventory was required for the two orders to execute. This elasticity separates a categorical market from a stock market, where the supply of shares is fixed in the short run and a concentrated exit must be absorbed by dealer inventory or a price concession.

This anatomy also shows the form that disagreement takes. Different interpretations of public news can generate large volume with little price movement \citep{kandelpearson1995}. At the withdrawal, disagreement took an asymmetric form: a small group crossed the book to cut Trump exposure, while dispersed wallets accommodated it through resting orders. Minting explains how these orders cleared, not why traders were willing to meet near the existing Trump YES price. The debate makes the distinction clear. Its Trump YES supply was also almost entirely minted, yet the news repriced Trump and the price rose.

\subsection{Why did the Trump YES price move so little?}
\label{sec:reallocation}

\begin{figure}[hbtp!]
	\centering
	\includegraphics[width=11cm]{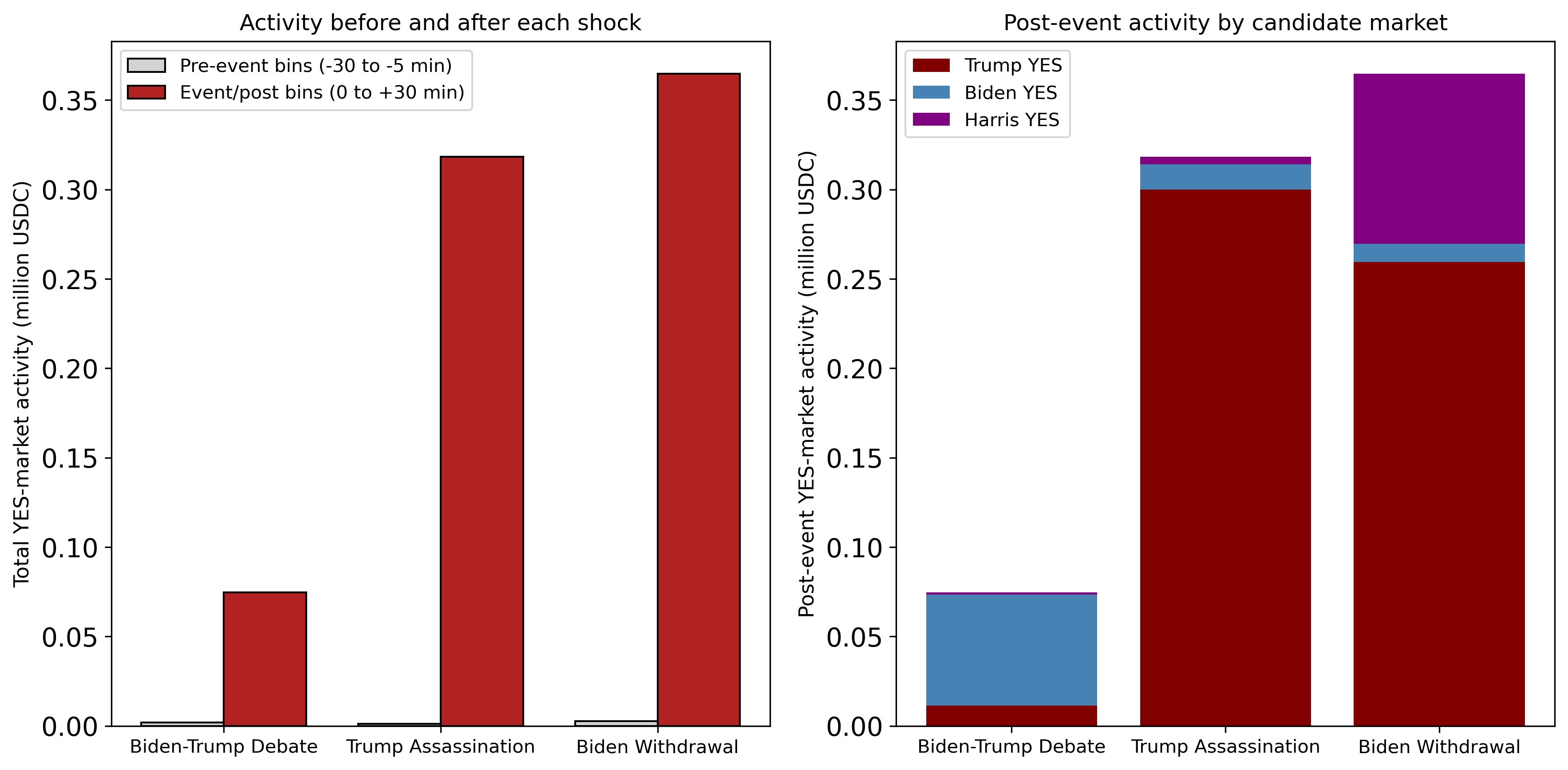}
	\caption{Total market activity by candidate market around each event}
	\label{fig_activity_decomp}
	\vspace{0.25em}
	\begin{minipage}{0.98\linewidth}
		\footnotesize \raggedright \textit{Notes:} The left panel compares pre-event bins $-6$ through $-1$ with event and post-event bins 0 through 6. The right panel decomposes activity in bins 0 through 6 across the Trump, Biden, and Harris YES markets. Activity equals trading volume plus the larger of YES minting and YES burning in each 5-minute bin, following \citet{tsang2026anatomy}, and is reported in millions of USDC.
	\end{minipage}
\end{figure}

The anatomy shows how the flow cleared. The harder question is why it cleared near the old price. One possible explanation is that the withdrawal, being Democratic news, generated little trading in Trump YES. \autoref{fig_activity_decomp} shows the opposite. The withdrawal is the largest of the three events by total activity across the candidate markets, and about 71\% of its post-event activity occurs in Trump YES. Trading reallocates from Biden toward Harris within the Democratic side, but it does not bypass Trump YES, so an absence of trading there cannot explain the muted response.

A second possibility is that the Trump market was unusually deep, so that a given order imbalance moved the price little. We measure effective depth with Kyle's $\lambda$, the sensitivity of the log-odds Trump price to signed order flow, where a larger $\lambda$ marks a thinner market in which flow moves the price more \citep{kyle1985continuous}. We aggregate initiator-signed flow into 5-minute bins and estimate
\begin{equation}
	\Delta \theta_t=\alpha+\lambda_t Q_t+\varepsilon_t,
\end{equation}
where $\theta_t$ is the log-odds Trump YES price and $Q_t$ is signed order flow in millions of USDC. At each point, the regression uses the trailing 2,016 bins, or seven days, with estimation details in \autoref{appendix_impact}. \autoref{fig_kyle_lambda} plots $\lambda_t$ across the event window. The withdrawal estimate remains above the debate and assassination estimates, so the Trump market was if anything less deep at the withdrawal, not more. Unusual depth cannot explain the flat price either.

The linked candidate prices point to the answer (\autoref{tab_candidate_price_changes}). During the withdrawal, the dominant repricing occurs within the Democratic field. In the first thirty minutes, Biden YES falls by about 7 cents and Harris YES rises by about 14 cents, while their combined probability rises by about 7 cents and Trump YES falls by about 3 cents. The 21-cent gross shift between Biden and Harris therefore dominates the movement in the Trump-versus-Democrat contest.

The comparison events sharpen the contrast. At the assassination attempt, Biden YES and Harris YES fall together as Trump YES rises, the signature of news about the Trump-versus-Democrat contest. The debate begins the Biden-to-Harris reallocation, with Biden YES down and Harris YES up. The sign of the Biden--Harris co-movement is the diagnostic: a common move marks news about the Trump-versus-Democrat axis that reprices Trump YES, while an offsetting move marks a Democratic reshuffle that leaves Trump YES comparatively stable. Price discovery follows the information content of order flow rather than its quantity \citep{hasbrouck1991measuring}. At the withdrawal, that content sat mainly in the Democratic candidate markets even though most activity occurred in Trump YES.

\subsection{Why did the withdrawal announcement carry little news?}
\label{sec:anticipation}

The reshuffle explains why the Trump YES price moved little, but a nominee switch could still have carried news about the eventual winner. A stronger or weaker replacement would change the chance that a Democrat wins, and with it the Trump price. The announcement carried little news because the market had anticipated the switch for weeks, an instance of the standard distinction between expected and unexpected news in event studies \citep{fama1969adjustment}. Most studies must infer anticipation indirectly. This setting measures it directly: Polymarket ran a \emph{Democratic Nominee 2024} market over the same period, so the probability that Biden would not remain the nominee is itself a traded price.

\begin{figure}[hbtp!]
	\centering
	\includegraphics[width=\textwidth]{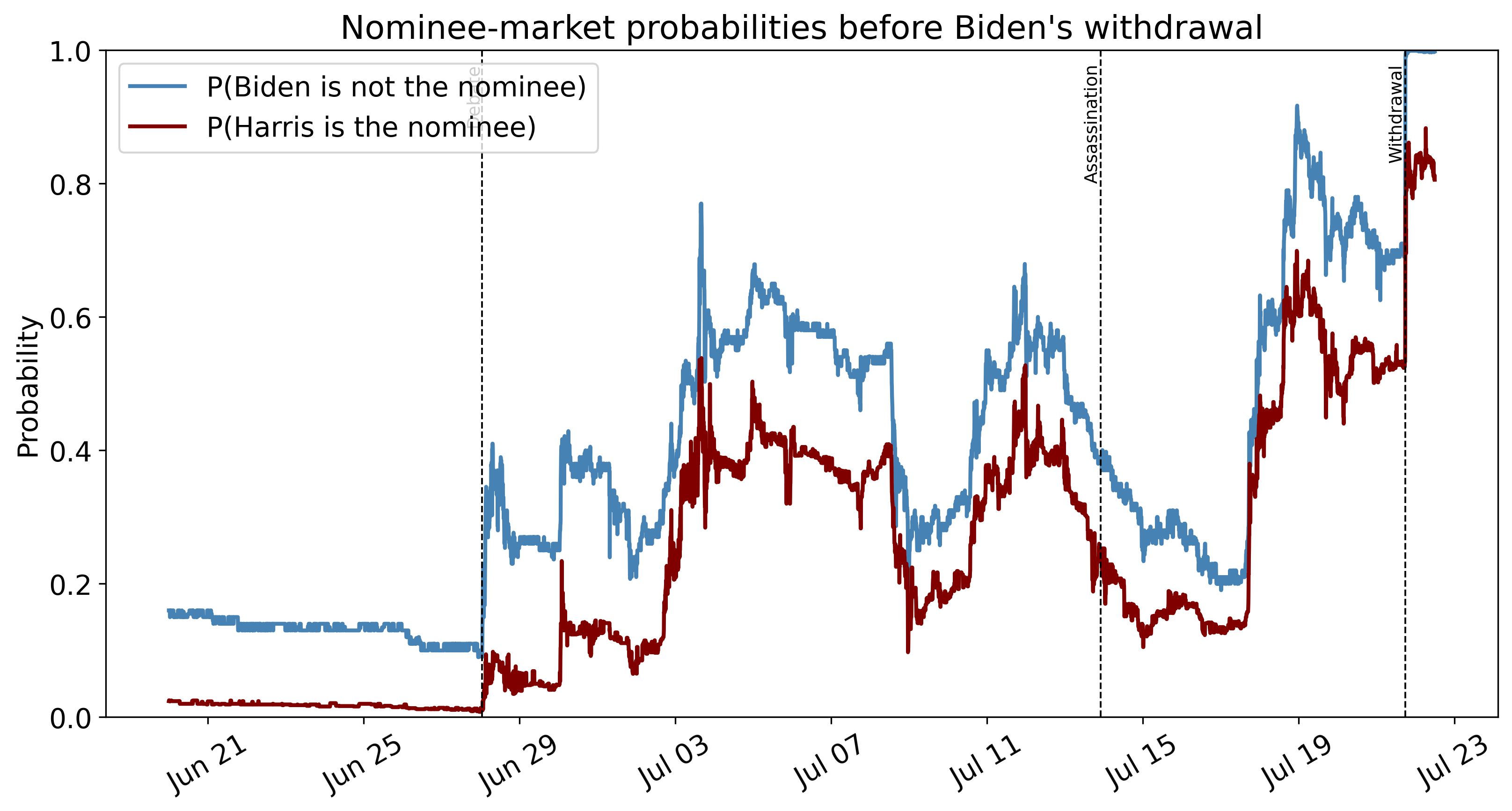}
	\caption{The market priced the switch before the announcement}
	\label{fig_anticipation}
	\vspace{0.25em}
	\begin{minipage}{0.98\linewidth}
	\footnotesize \raggedright \textit{Notes:} The blue line maps Biden-nominee YES and NO prices into the probability that Biden is not the nominee and combines them within 5-minute bins using activity weights. The maroon line analogously reports the Harris-nominee probability. Key levels are reported in \autoref{tab_kuttner}, Panel A.
	\end{minipage}
\end{figure}

The nominee market shows how much of the switch was anticipated (\autoref{fig_anticipation}). We measure the switch probability as one minus the Biden-nominee price. It stood at 0.10 on the eve of the debate and reached 0.65 by July 5. It drifted back to 0.38 before the assassination attempt, fell to 0.29 afterward, and recovered to 0.70 one minute before the withdrawal announcement (\autoref{tab_kuttner}, Panel A). Much of the switch had therefore been priced before the announcement.

Following the surprise-decomposition logic of \citet{kuttner2001monetary}, the nominee market also lets us calibrate the remaining surprise. Writing the Trump YES price as the probability-weighted average of its value under a switch and under no switch gives
\begin{equation}
\Delta p^{T}_{\text{ann}} \;\approx\; \big(1-P_{\text{ant}}\big)\times\big[\,\mathbb{E}(p^{T}\,|\,\text{switch})-\mathbb{E}(p^{T}\,|\,\text{no switch})\,\big],
\end{equation}
where $P_{\text{ant}}$ is the pre-announcement switch probability and the bracketed term is the conditional Trump YES price gap. We infer that gap from pre-announcement co-movement between the Trump YES price and the switch probability. Because the estimated slope is sensitive to sampling frequency, we report hourly and daily estimates, and together they imply an announcement move between $-1.0$ and $-3.1$ cents. Using the exact announcement-time window, the realized thirty-minute move was $-2.6$ cents, essentially unchanged at three hours (\autoref{tab_kuttner}, Panel B). The observed anticipation therefore accounts quantitatively for the small withdrawal response. The debate provides a useful contrast. The switch channel predicts a Trump move between $-0.8$ and $-2.2$ cents, while the realized three-hour move was $+1.7$ cents (Panel C). The opposite sign shows that the debate contained Trump-versus-Democrat news beyond the anticipated-switch channel.

\begin{figure}[hbtp!]
	\centering
	\includegraphics[width=\textwidth]{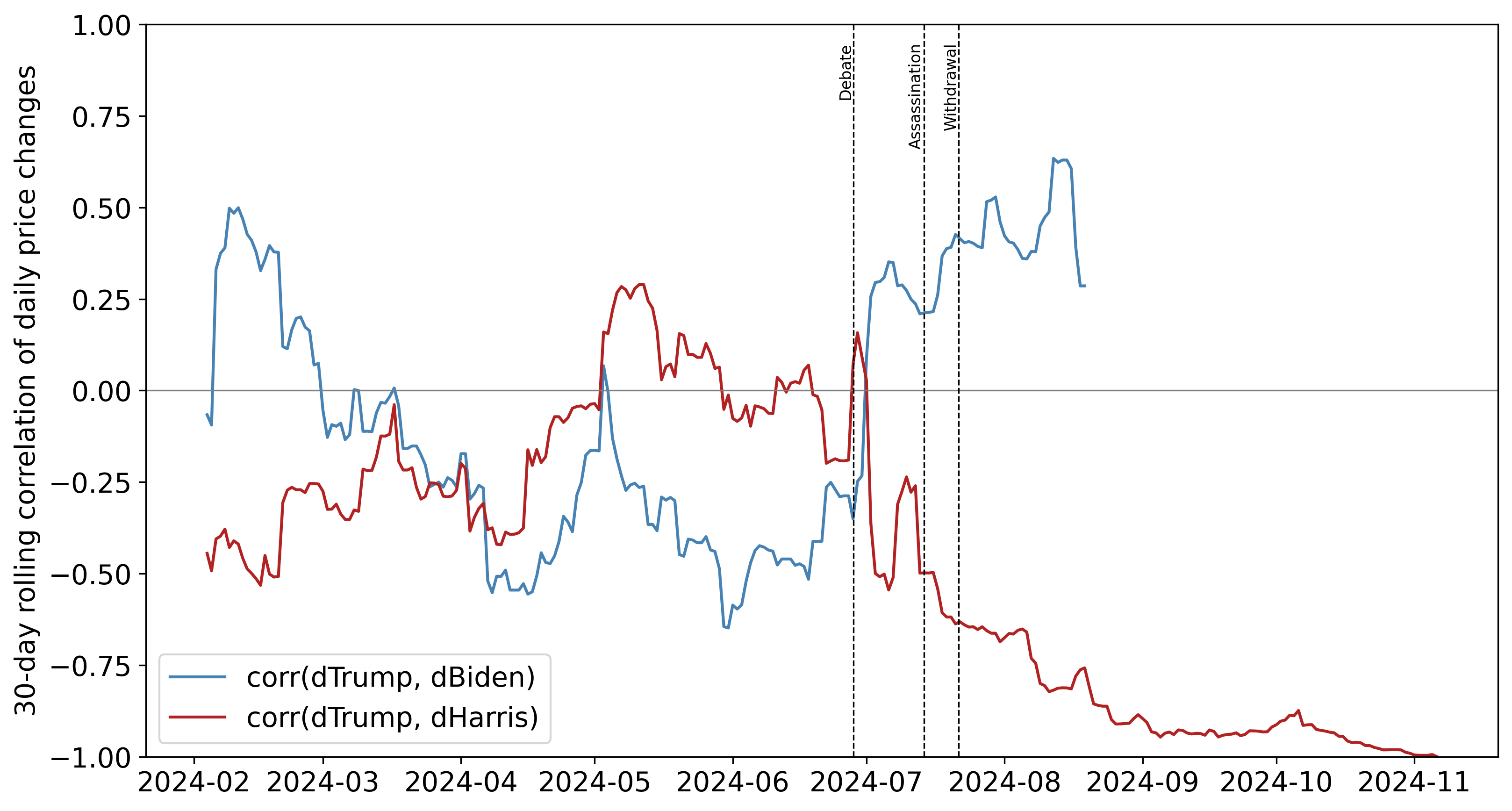}
	\caption{Rolling correlations among candidate YES prices}
	\label{fig_rollcorr_candidates}
	\vspace{0.25em}
	\begin{minipage}{0.98\linewidth}
	\footnotesize \raggedright \textit{Notes:} Thirty-day rolling correlations pair daily Trump YES closing-price changes with Biden YES (blue) and Harris YES (red). Vertical lines mark the three shocks.
	\end{minipage}
\end{figure}

If traders had already priced the switch, the Trump price should have been tracking the likely replacement before the announcement. The co-movement between the Trump YES price and the two Democratic candidates' YES prices shows exactly this pattern. \autoref{fig_rollcorr_candidates} plots thirty-day rolling correlations of daily Trump YES changes with the Biden and Harris YES changes. In the months before the debate, the Trump change is negatively correlated with the Biden change, the signature of a Trump--Biden contest. After the debate, the correlations rotate: the Trump--Harris correlation turns strongly negative while the Trump--Biden correlation turns positive, weeks before the announcement. The Trump price was already tracking the likely replacement rather than the departing incumbent.

\subsection{Why did the price reverse or persist?}
\label{sec:dynamics}

\begin{figure}[hbtp!]
	\centering
	\includegraphics[width=\textwidth]{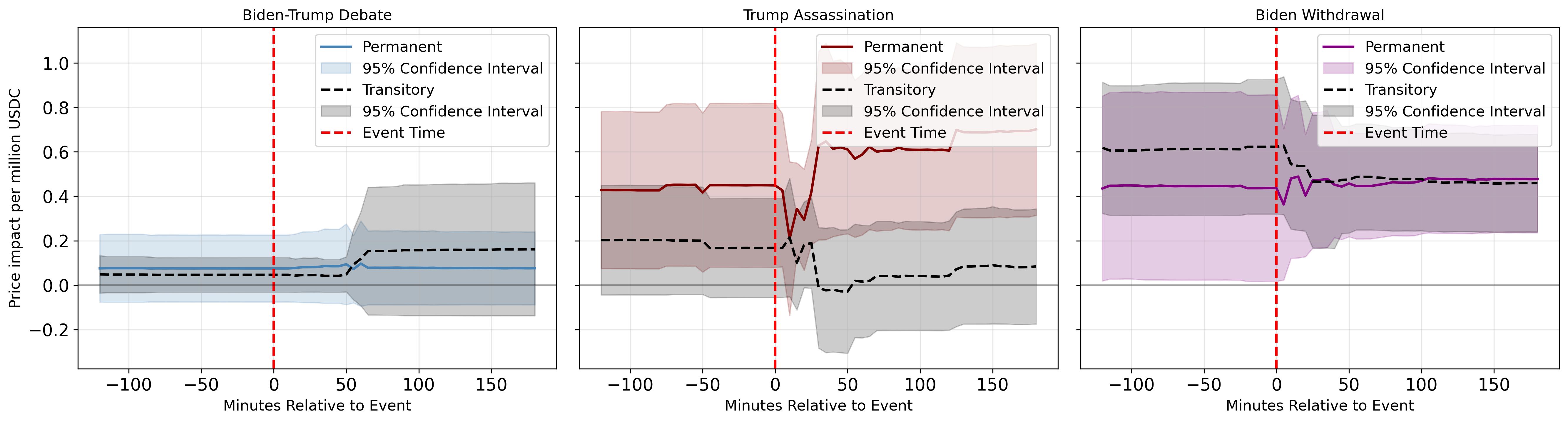}
	\caption{Rolling permanent and transitory price impact around political shocks}
	\label{fig_glosten_harris}
	\vspace{0.25em}
	\begin{minipage}{0.98\linewidth}
	\footnotesize \raggedright \textit{Notes:} Rolling estimates use $\Delta\theta_t=\alpha+\lambda_{\mathrm{perm},t}Q_t+\lambda_{\mathrm{trans},t}\Delta Q_t+\varepsilon_t$, where $\theta_t$ is the Trump YES log-odds price and $Q_t$ is initiator-signed flow in millions of USDC. Each uses 2,016 trailing five-minute bins and HAC(12) standard errors. Solid and dashed lines show permanent and transitory components. Bands are 95\% confidence intervals.
	\end{minipage}
\end{figure}

The reallocation and anticipation evidence explains why the withdrawal barely moves the Trump price. It leaves open the final question: why the debate's initial increase reverses while the assassination response persists. By three hours, about half of the debate's initial thirty-minute increase remains, whereas the assassination increase has almost doubled. We separate more permanent repricing from temporary pressure with the specification of \citet{glostenharris1988}:
\begin{equation}
	\Delta \theta_t=\alpha+\lambda_{\mathrm{perm},t}Q_t
	+\lambda_{\mathrm{trans},t}\Delta Q_t+\varepsilon_t,
\end{equation}
in which the coefficient on signed order flow $Q_t$ is the more permanent component and the coefficient on its change $\Delta Q_t$ is the transitory component. We sign flow from the initiating order identified by the ledger and estimate the regression at each 5-minute point over the same trailing seven-day window used for Kyle's $\lambda$. \autoref{fig_glosten_harris} displays the coefficients over the $[-2,+3]$ hour event window.

The rolling estimates track the observed paths. After the debate, the transitory component rises above the permanent component as the initial Trump increase unwinds. After the assassination attempt, the permanent component dominates and the transitory component remains near zero, consistent with a move to a new price level. At the withdrawal, the two components remain close together. The decomposition therefore summarizes the difference between the debate's reversal and the assassination response's persistence, while the linked candidate markets identify the news content behind those price paths.

Taken together, the three paths have one explanation: the Trump YES price followed the news content about the Trump-versus-Democrat contest and how much of it was anticipated, not the volume of trading. The debate and the assassination attempt both carried Trump-relevant news, but the debate's initial move included a large transitory component that unwound, while the assassination repricing was permanent. The withdrawal generated the heaviest flow and the least news: the switch was largely anticipated, the repricing stayed within the Democratic field, and the flow cleared through dispersed resting orders and newly minted shares.

\section{Conclusion}
\label{sec:conclusion}

\noindent Using the complete wallet-resolved ledger of Polymarket's 2024 U.S. presidential election market, we study three political shocks that produced reversal, persistence, and little movement in the Trump YES price. The ledger connects which wallets supply the immediate trading response to how orders clear and prices adjust across linked candidate markets.

On the trader side, pre-event activity is the most stable correlate of post-shock trading. New entry rises at the assassination attempt and withdrawal but not at the debate. Incumbents with greater prior volume increase their activity more after every shock, most strongly at the assassination attempt and withdrawal. At those two events, realized portfolio gains are also associated with larger increases in activity, while the debate response is stronger on the loss side. Active incumbents thus supply the trading response at every shock, while realized exposure helps explain which side becomes more active in each case.

Prices follow a different logic. At Biden's withdrawal, a concentrated group cut Trump exposure against dispersed resting demand, and most of the Trump YES shares supplied to buyers were newly minted. The ledger therefore explains how heavy flow cleared near the existing price. The linked markets explain why traders met there: probability shifted mainly from Biden YES to Harris YES, and nominee-market prices show that much of the switch was anticipated before the announcement. The Trump YES price therefore moved little even though the event revalued existing portfolios and brought traders back. Across the comparison events, the rolling price-impact decomposition tracks the debate's reversal with a stronger transitory component and the assassination response's persistence with a stronger permanent component.

Together, the results show that heavy trading and a large price revision need not coincide, and why the two separate so readily in a categorical market. Disagreement can generate volume without a price move in a stock market too, but a fixed short-run share supply forces a concentrated exit through dealer inventory or a price concession. A categorical market can instead mint new YES--NO pairs when buyers stand on both sides, or burn pairs when sellers do, so heavy flow clears without dealer inventory or a price concession. Prior activity and realized portfolio exposure help explain which traders return after a shock. A candidate's YES price instead depends on what the news reveals about that candidate and how much was already priced. Volume is therefore evidence that traders responded, not sufficient evidence of belief revision. The evidence comes from three shocks in one market, and we observe executed trades rather than the standing order book. Future work can test the same distinction between participation and belief revision in other categorical markets.

\clearpage

\bibliographystyle{plainnat}
\bibliography{ref}

\clearpage


\begin{table}[!htbp]
	\centering
	\caption{Measurement frequency by variable}
	\label{tab_measurement_map}
	\begin{threeparttable}
	\footnotesize
	\begin{tabularx}{\textwidth}{p{0.5\textwidth}>{\raggedright\arraybackslash}X}
	\toprule
	Variable & Frequency\\
\midrule
New trader entry & 5-min event time \\
Abnormal incumbent activity & 5-min event time \\
Trader activity characteristics & prior 30 days, ending 30 min before event \\
Log-odds prices and returns & 5-min event time \\
Order-level direction and initiator-signed flow & 5-min event time \\
Kyle $\lambda$ and permanent-transitory decomposition & rolling 5-min estimates over 7-day windows \\
Switch probability (Democratic-nominee market) & 5-min event time \\
	\bottomrule
\end{tabularx}
\begin{tablenotes}[para,flushleft]
\item
	\footnotesize \raggedright \textit{Notes:} Event-time variables are assigned to 5-minute bins using the ceiling convention. The same event window applies to all three shocks. The three-hour horizon has a separate role, describing whether the initial price response persists or reverses.
\end{tablenotes}
\end{threeparttable}
\end{table}

\begin{table}[!htbp]
\caption{Trader characteristics}
\setlength\tabcolsep{1.5pt} 
\begin{threeparttable}
\footnotesize
\begin{tabularx}{\textwidth}{l|>{\raggedright\arraybackslash}X}
\toprule
Characteristics & Description \\
\hline
	Trad Vol High & equals 1 if the trader's prior-month volume is in the top 50\% among active traders \\
		Trad Freq High & equals 1 if the trader traded more than once during the prior month \\
	Trad Int Multi & equals 1 if the trader trades in multiple 5-minute intervals during the pre-event measurement window \\
	MtM Loss & equals 1 if the trader's event mark-to-market revaluation is below $-1$ USDC \\
	MtM Gain & equals 1 if the trader's event mark-to-market revaluation is above $1$ USDC \\
Single Market & equals 1 if the trader only trades in one market \\
Contrarian & equals 1 if the trader is a contrarian trader (classified before the first event) \\
Momentum & equals 1 if the trader is a momentum trader (classified before the first event) \\
\bottomrule
\end{tabularx}  
\begin{tablenotes}[para,flushleft]
	\item
	\footnotesize \raggedright \textit{Notes:} Prior activity is measured before each event. Holdings used in MtM are fixed before the event window, and the common realized price vector is measured over the first thirty post-event minutes. Contrarian and momentum types are classified before the first event and held fixed.
\end{tablenotes}
\end{threeparttable}
\label{table_trader_characteristics}
\end{table}

\begin{table}[!htbp]
\caption{Trader characteristics: summary statistics}
\setlength\tabcolsep{1.5pt}
\begin{threeparttable}
\begin{tabular}{l|cccc}
	\toprule
	Characteristics & Full Sample & Biden--Trump Debate & Trump Assassination & Biden Withdrawal \\
	\hline
	Trad Vol High & 31.03\% & 30.83\% & 29.16\% & 32.61\% \\
	Trad Freq High & 23.98\% & 21.99\% & 21.10\% & 27.41\% \\
	Trad Int Multi & 19.70\% & 17.69\% & 17.77\% & 22.40\% \\
	MtM Loss & 27.69\% & 18.87\% & 22.32\% & 37.16\% \\
	MtM Gain & 24.53\% & 27.89\% & 33.61\% & 15.45\% \\
	Single Market & 89.64\% & 90.23\% & 89.15\% & 89.67\% \\
	Contrarian & 1.55\% & 3.25\% & 1.24\% & 0.78\% \\
	Momentum & 1.49\% & 3.34\% & 1.03\% & 0.75\% \\
	\bottomrule
\end{tabular}
	\begin{tablenotes}[para,flushleft]
	\item
	\footnotesize \raggedright \textit{Notes:} Entries are sample shares. The full sample pools the three events.
	\end{tablenotes}
\end{threeparttable}
	\label{table_summary_trader_characteristics}
\end{table}

\begin{landscape}
\begin{table}[p]
\centering
\caption{Stacked panel regressions: trading volume}
\label{tab_panel_volume_5min}
\begin{threeparttable}
\small
\setlength{\tabcolsep}{5pt}
\begin{tabular}{lccccccc}
	\toprule
	& (1) & (2) & (3) & (4) & (5) & (6) & (7) \\
	& Baseline & Trading Vol & Trading Freq & MtM Exposure & Single Market & Trader Types & Full \\
	\midrule
	Post Event & 0.0230*** & & & & & & \\
	& (0.0012) & & & & & & \\
	Post Event $\times$ Trad Vol High & & 0.0688*** & & & & & 0.0373*** \\
	& & (0.0040) & & & & & (0.0035) \\
	Post Event $\times$ Trad Freq High & & & 0.0705*** & & & & 0.0480*** \\
	& & & (0.0048) & & & & (0.0050) \\
	Post Event $\times$ MtM Loss & & & & 0.0230*** & & & -0.0034 \\
	& & & & (0.0025) & & & (0.0026) \\
	Post Event $\times$ MtM Gain & & & & 0.0639*** & & & 0.0417*** \\
	& & & & (0.0044) & & & (0.0034) \\
	Post Event $\times$ Single Mkt & & & & & -0.0093* & & 0.0047 \\
	& & & & & (0.0055) & & (0.0058) \\
	Post Event $\times$ Contrarian & & & & & & 0.0325** & -0.0147 \\
	& & & & & & (0.0157) & (0.0160) \\
	Post Event $\times$ Momentum & & & & & & 0.0763*** & 0.0273 \\
	& & & & & & (0.0243) & (0.0245) \\
	\midrule
	Entity Effects & YES & YES & YES & YES & YES & YES & YES \\
	Time Effects & NO & YES & YES & YES & YES & YES & YES \\
	R-Squared (Within) & 0.0016 & 0.0047 & 0.0042 & 0.0036 & -0.0014 & 0.0005 & 0.0059 \\
	R-Squared (Between) & 0.0185 & 0.0499 & 0.0453 & 0.0378 & -0.0130 & 0.0055 & 0.0657 \\
	R-Squared (Overall) & 0.0036 & 0.0100 & 0.0091 & 0.0077 & -0.0027 & 0.0011 & 0.0130 \\
	Observations & 284420 & 284420 & 284420 & 284420 & 284420 & 284420 & 284420 \\
	\bottomrule
\end{tabular}
	\begin{tablenotes}[para,flushleft]
	\item
	\footnotesize \raggedright \textit{Notes:} The dependent variable is trader-level $\operatorname{asinh}(\text{trading volume})$ in 5-minute event time. Post Event equals one at and after the shock. Regressors interact Post Event with the characteristics in \autoref{table_trader_characteristics}. Entity effects are trader-by-event fixed effects. Time effects are event-time-bin fixed effects. Standard errors, clustered by trader-event, are in parentheses. * p$<$0.1, ** p$<$0.05, *** p$<$0.01.
	\end{tablenotes}
\end{threeparttable}
\end{table}
\end{landscape}

\begin{landscape}
\begin{table}[p]
\centering
\caption{Stacked panel regressions: trading frequency}
\label{tab_panel_freq_5min}
\begin{threeparttable}
\small
\setlength{\tabcolsep}{5pt}
\begin{tabular}{lccccccc}
	\toprule
	& (1) & (2) & (3) & (4) & (5) & (6) & (7) \\
	& Baseline & Trading Vol & Trading Freq & MtM Exposure & Single Market & Trader Types & Full \\
	\midrule
	Post Event & 0.0041*** & & & & & & \\
	& (0.0002) & & & & & & \\
	Post Event $\times$ Trad Vol High & & 0.0117*** & & & & & 0.0057*** \\
	& & (0.0008) & & & & & (0.0007) \\
	Post Event $\times$ Trad Freq High & & & 0.0130*** & & & & 0.0092*** \\
	& & & (0.0009) & & & & (0.0010) \\
	Post Event $\times$ MtM Loss & & & & 0.0047*** & & & 0.0004 \\
	& & & & (0.0005) & & & (0.0006) \\
	Post Event $\times$ MtM Gain & & & & 0.0100*** & & & 0.0064*** \\
	& & & & (0.0008) & & & (0.0006) \\
	Post Event $\times$ Single Mkt & & & & & -0.0029** & & -0.0003 \\
	& & & & & (0.0013) & & (0.0013) \\
	Post Event $\times$ Contrarian & & & & & & 0.0050* & -0.0035 \\
	& & & & & & (0.0026) & (0.0028) \\
	Post Event $\times$ Momentum & & & & & & 0.0149*** & 0.0061 \\
	& & & & & & (0.0051) & (0.0052) \\
	\midrule
	Entity Effects & YES & YES & YES & YES & YES & YES & YES \\
	Time Effects & NO & YES & YES & YES & YES & YES & YES \\
	R-Squared (Within) & 0.0016 & 0.0043 & 0.0043 & 0.0031 & -0.0026 & 0.0006 & 0.0058 \\
	R-Squared (Between) & 0.0167 & 0.0407 & 0.0411 & 0.0292 & -0.0217 & 0.0049 & 0.0562 \\
	R-Squared (Overall) & 0.0036 & 0.0091 & 0.0092 & 0.0066 & -0.0051 & 0.0011 & 0.0125 \\
	Observations & 284420 & 284420 & 284420 & 284420 & 284420 & 284420 & 284420 \\
	\bottomrule
\end{tabular}
	\begin{tablenotes}[para,flushleft]
	\item
	\footnotesize \raggedright \textit{Notes:} The dependent variable is trader-level $\operatorname{asinh}(\text{trading frequency})$ in 5-minute event time. Post Event equals one at and after the shock. Regressors interact Post Event with the characteristics in \autoref{table_trader_characteristics}. Entity effects are trader-by-event fixed effects. Time effects are event-time-bin fixed effects. Standard errors, clustered by trader-event, are in parentheses. * p$<$0.1, ** p$<$0.05, *** p$<$0.01.
	\end{tablenotes}
\end{threeparttable}
\end{table}
\end{landscape}

\begin{table}[hbtp!]
	\centering
	\caption{Order-level anatomy of the withdrawal clearing, $[0,+30\text{ min})$}
	\label{tab_clearing_anatomy}
	\begin{threeparttable}
	\footnotesize
	\begin{tabular}{lr}
		\toprule
		\multicolumn{2}{l}{\emph{Panel A: taker-order flow by leg}}\\
		\midrule
		Sell Trump YES: USDC (wallets) & \$160,615 \; (13) \\
		Buy Trump NO: USDC (wallets) & \$102,108 \; (16) \\
		Buy Trump YES: USDC (wallets) & \$13,375 \; (27) \\
		Sell Trump NO: USDC (wallets) & \$2,154 \; (2) \\
		Share of taker-order flow reducing Trump exposure & 0.944 \\
		\midrule
		\multicolumn{2}{l}{\emph{Panel B: all wallets, classified by net exposure change}}\\
		\midrule
		Wallets raising exposure: count (net USDC) & 261 \; ($+$\$231,505) \\
		\quad share of volume executed as taker orders & 0.138 \\
		Wallets reducing exposure: count (net USDC) & 37 \; ($-$\$227,245) \\
		\quad share of volume executed as taker orders & 0.892 \\
		\quad top-5 wallets' share of reducing flow & 0.870 \\
		\bottomrule
	\end{tabular}
	\begin{tablenotes}[para,flushleft]
		\footnotesize
		\item \textit{Notes:} Panel A signs each match by its ledger-identified taker order (\autoref{appendix_trade_direction}). Panel B classifies wallets by their net Trump-exposure change across all YES and NO fills.
	\end{tablenotes}
	\end{threeparttable}
\end{table}

\begin{table}[hbtp!]
	\centering
	\caption{Fill mechanism and wallet counts in the Trump YES market}
	\label{tab_fill_mechanism}
	\vspace{0.25em}
	\begin{tabular}{lccc}
		\toprule
		 & Debate & Assassination & Withdrawal \\
		\midrule
		Minted share of Trump YES supply & 0.99 & 0.14 & 0.66 \\
		Wallets raising Trump exposure & 12 & 111 & 261 \\
		Wallets absorbing Trump exposure & 12 & 139 & 37 \\
		\bottomrule
	\end{tabular}
	\vspace{0.25em}
	\begin{minipage}{0.92\linewidth}
	\footnotesize \raggedright \textit{Notes:} Quantities use raw ledger timestamps from each shock up to 30 minutes afterward. Minted share is new Trump YES divided by Trump YES supplied to buyers. Raising (absorbing) wallets increase (decrease) net Trump exposure across YES and NO fills.
	\end{minipage}
\end{table}

\begin{table}[hbtp!]
	\centering
	\caption{Candidate YES-price changes at 30 minutes and three hours}
	\label{tab_candidate_price_changes}
	\begin{threeparttable}
	\footnotesize
	\begin{tabular}{llrrrr}
		\toprule
		Event & Horizon & $\Delta p^{T}$ & $\Delta p^{B}$ & $\Delta p^{H}$ & $\Delta(p^{B}+p^{H})$ \\
		\midrule
		Biden--Trump Debate & 30 minutes & $+0.04$ & $-0.05$ & $+0.01$ & $-0.03$ \\
		 & 3 hours & $+0.02$ & $-0.13$ & $+0.03$ & $-0.10$ \\
		Trump Assassination & 30 minutes & $+0.05$ & $-0.04$ & $-0.02$ & $-0.06$ \\
		 & 3 hours & $+0.10$ & $-0.05$ & $-0.05$ & $-0.10$ \\
		Biden Withdrawal & 30 minutes & $-0.03$ & $-0.07$ & $+0.14$ & $+0.07$ \\
		 & 3 hours & $-0.01$ & $-0.07$ & $+0.12$ & $+0.05$ \\
		\bottomrule
	\end{tabular}
	\begin{tablenotes}[para,flushleft]
		\footnotesize
		\item \textit{Notes:} Changes run from the last complete 5-minute pre-event price bin to each horizon. The final column sums Biden and Harris. Omitted minor candidates prevent the columns from summing to zero.
	\end{tablenotes}
	\end{threeparttable}
\end{table}

\begin{table}[!htbp]
	\centering
	\caption{Anticipation and the Kuttner surprise decomposition at the withdrawal}
	\label{tab_kuttner}
	\begin{threeparttable}
	\footnotesize
	\begin{tabular}{lr}
		\toprule
		\multicolumn{2}{l}{\emph{Panel A: switch probability $P(\text{Biden not nominee})$ from the nominee market}}\\
		\midrule
		Debate eve (Jun 27) & 0.100 \\
		One day after the debate & 0.260 \\
		July 5 & 0.650 \\
		One hour before the assassination attempt & 0.383 \\
		One day after the assassination attempt & 0.290 \\
		One day before the withdrawal & 0.740 \\
		One minute before the withdrawal announcement & 0.702 \\
		Thirty minutes after & 0.992 \\
		\midrule
		\multicolumn{2}{l}{\emph{Panel B: surprise decomposition of the Trump YES price move}}\\
		\midrule
		Conditional Trump YES price gap, hourly co-movement (s.e.) & $-0.035$ \; (0.019) \\
		Conditional Trump YES price gap, daily co-movement (s.e.) & $-0.103$ \; (0.045) \\
		Predicted announcement move $(1-0.702)\times\text{gap}$ & $-0.010$ to $-0.031$ \\
		Realized Trump YES price move, 30 minutes & $-0.026$ \\
		Realized Trump YES price move, 3 hours & $-0.027$ \\
		\midrule
		\multicolumn{2}{l}{\emph{Panel C: the debate through the same lens}}\\
		\midrule
		Debate change in switch probability & $+0.216$ \\
		Predicted Trump YES price move through the switch channel & $-0.008$ to $-0.022$ \\
		Realized debate Trump YES price move, 3 hours & $+0.017$ \\
		\bottomrule
	\end{tabular}
	\begin{tablenotes}[para,flushleft]
		\footnotesize
		\item \textit{Notes:} Switch probability maps Biden-nominee YES and NO prices to the Biden-not-nominee payoff and combines them within 5-minute bins using activity weights. Conditional gaps are HAC slopes of Trump YES changes on switch-probability changes. The hourly sample runs from the debate through one hour before withdrawal ($N=564$), and the daily sample uses June 28--July 20 ($N=23$).
	\end{tablenotes}
	\end{threeparttable}
\end{table}

\clearpage

\begin{figure}[hbtp!]
	\centering
	\includegraphics[width=\textwidth]{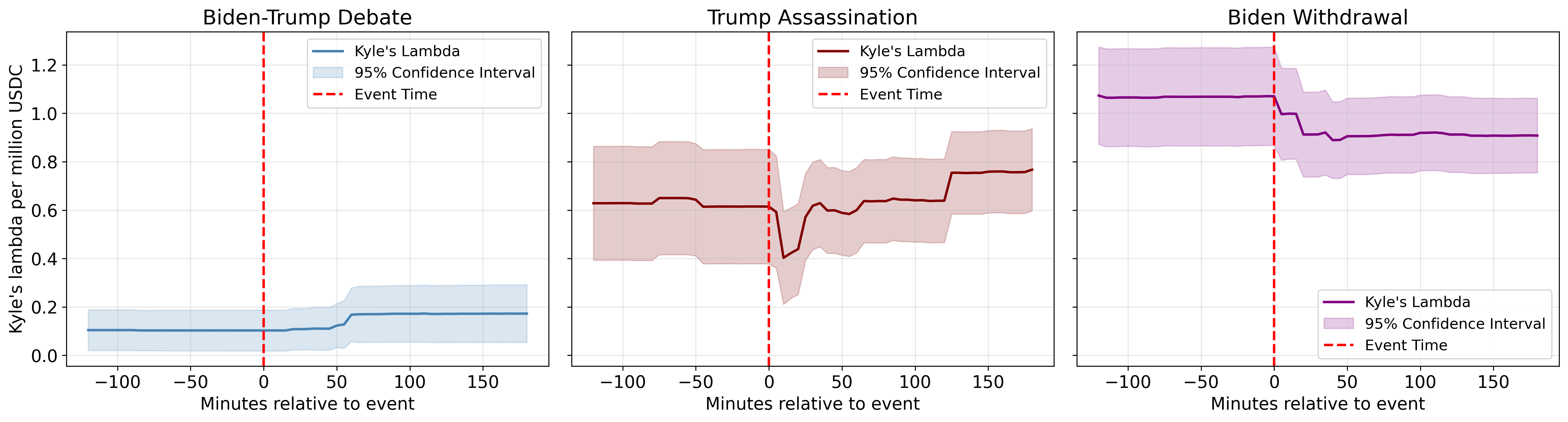}
	\caption{Event-time Kyle's $\lambda$ for the Trump YES market}
	\label{fig_kyle_lambda}
	\vspace{0.25em}
	\begin{minipage}{0.98\linewidth}
	\footnotesize \raggedright \textit{Notes:} Panels plot rolling Kyle-style price impact from two hours before to three hours after each event. Signed order flow is read from the initiating order identified by the ledger, with purchases of Trump YES and sales of Trump NO signed positively. At each 5-minute point, $\lambda$ is estimated over the trailing 2,016 bins. Coefficients are scaled per million USDC. Shaded areas are 95\% confidence intervals. Larger values indicate greater price impact and lower effective depth.
	\end{minipage}
\end{figure}

\clearpage

\appendix
\counterwithin{figure}{section}
\counterwithin{table}{section}
\setcounter{page}{1}

\begin{center}
{\Large\bfseries Online Appendix}
\end{center}

\noindent \autoref{appendix_trade_direction} explains how the ledger identifies order direction and the initiating side and validates the resulting flow measure. \autoref{appendix_event_study} defines abnormal incumbent activity and reports matched weekday-time placebo tests. \autoref{appendix_impact} presents the estimation details for the rolling Kyle and permanent--transitory price-impact measures. \autoref{appendix_robustness} reports the pooled participation specification, which avoids counting multiple fills generated by one order. \autoref{appendix_event_specific_panel} reports the volume, frequency, and participation regressions separately for each shock.


\section{Reading Trade Direction from the Ledger}
\label{appendix_trade_direction}

\noindent Polymarket matches orders off chain and settles each match on chain. Each filled order produces an \texttt{OrderFilled} event that records the order's signer in the \texttt{maker} field and the assets that the signer gives and receives. USDC has asset ID 0. If \texttt{makerAssetId}=0, the signer pays USDC and buys the share in \texttt{takerAssetId}. If \texttt{takerAssetId}=0, the signer delivers the share in \texttt{makerAssetId} and sells it. Buying YES or selling NO raises exposure to the candidate winning, while selling YES or buying NO lowers it. Wallet-level attribution therefore uses the \texttt{maker} field of each \texttt{OrderFilled} event.

The ledger also identifies which order initiated the match. The exchange contract's \texttt{matchOrders} function receives one active taker order and one or more resting maker orders.\footnote{Polymarket's CLOB documentation defines the taker as the order that crosses the spread and executes against resting orders. The Polymarket CTF Exchange source (\texttt{Trading.sol} and \texttt{CTFExchange.sol}) describes the taker order as the active order supplied to \texttt{matchOrders}.} A resting order's fill names the incoming wallet in its \texttt{taker} field. The incoming order's own fill instead has the exchange contract in its \texttt{taker} field, so the \texttt{maker} on that row is the initiating wallet. The transaction-level \texttt{OrdersMatched} event repeats the same wallet in \texttt{takerOrderMaker}. These fields identify the initiator without reconstructing the unobserved order book.

\autoref{tab_fill_routes} illustrates the rule with two matches. Panel A contains one resting order and one incoming order. Panel B shows a mixed match in which one incoming Trump YES buy is filled partly by a resting YES seller and partly by a resting NO buyer through share minting. In both panels, the row whose \texttt{taker} is the exchange contract identifies the incoming wallet, and the \texttt{OrdersMatched} row repeats that wallet. The USDC leg then determines whether each wallet buys or sells. The mixed example shows that the same reading applies when one incoming order meets several resting orders through different fill routes.

Ledger-wide checks validate the classification rule. Across all 3,654,974 matches in the three candidate markets, each transaction has exactly one \texttt{OrdersMatched} event and one \texttt{OrderFilled} row whose \texttt{taker} is the exchange contract. The exchange contract never appears as a \texttt{maker}, and the exchange-row wallet, the non-exchange \texttt{taker} named on the resting-order rows, and \texttt{OrdersMatched.takerOrderMaker} agree in every transaction. The designated taker's direction also agrees with the price change in 97.9\% of the 747,451 transactions with a price change. Orders that demonstrably rested before later appearing in the taker slot account for only 0.65\% of taker volume and none of the volume in the three event windows.

As a final check, we compare ledger signing with the tick rule of \citet{lee1991inferring} on the same transactions. The two methods agree on 96.8\% of transactions and 92.8\% of volume. Among disagreements, 69.7\% occur at zero ticks, where the tick rule carries forward a previous sign. \autoref{tab_dual_signing} shows that every event-window net-flow sign is unchanged. The paper therefore uses the initiator recorded in the ledger, while the tick rule provides an independent robustness check.

\section{Construction of Abnormal Incumbent Activity}
\label{appendix_event_study}

\noindent Following \citet{brown1985using} and \citet{mackinlay1997event}, we index 5-minute bins relative to event $e$ by $k$. The event window is $\mathcal{K}^{\mathrm{evt}}=\{-6,\ldots,6\}$, or thirty minutes before through thirty minutes after the event. The nonoverlapping estimation window is $\mathcal{K}^{\mathrm{est}}=\{-42,\ldots,-7\}$, the preceding 36 bins. For incumbent trader $i$, let $V_{iek}$ denote trading volume, $F_{iek}$ trading frequency, and $D_{iek}=\mathbbm{1}[V_{iek}>0]$ participation. The trader-event baselines are
\begin{align}
	\bar V_{ie}&=\frac{1}{36}\sum_{k\in\mathcal{K}^{\mathrm{est}}}V_{iek}, &
	\bar F_{ie}&=\frac{1}{36}\sum_{k\in\mathcal{K}^{\mathrm{est}}}F_{iek}, &
	\bar D_{ie}&=\frac{1}{36}\sum_{k\in\mathcal{K}^{\mathrm{est}}}D_{iek}.
\end{align}
We define the bin-level abnormal outcomes as
\begin{align}
	AV_{iek}&=\operatorname{asinh}(V_{iek})-\operatorname{asinh}(\bar V_{ie}),\\
	AF_{iek}&=\operatorname{asinh}(F_{iek})-\operatorname{asinh}(\bar F_{ie}),\\
	AP_{iek}&=D_{iek}-\bar D_{ie}.
\end{align}
For the volume series, we set $AV_{iek}=0$ when $V_{iek}=0$. This convention assigns the no-trade margin to the participation outcome rather than treating a zero-volume bin as negative abnormal volume.

For outcome $Y\in\{AV,AF,AP\}$, we first average across the $N_e$ incumbent traders in the event sample and then cumulate from the start of the event window:
\begin{equation}
	\widehat\mu^Y_{ek}=\frac{1}{N_e}\sum_iY_{iek},
	\qquad
	CAA^Y_{e\tau}=\sum_{k=-6}^{\tau}\widehat\mu^Y_{ek}.
\end{equation}
The bin-level standard error is the cross-sectional standard deviation of $Y_{iek}$ divided by $\sqrt{N_e}$. The displayed 95\% confidence bands cumulate those standard errors as $\sqrt{\sum_{k=-6}^{\tau}(\widehat{se}^{Y}_{ek})^2}$, which assumes independence across 5-minute bins.

We assess whether the observed increases are unusual with matched weekday-time placebos. For each event, we draw 500 pseudo-event timestamps with replacement from 5-minute bins that share the event's weekday and clock time, excluding timestamps within 24 hours of any real event. The test statistic is mean activity in bins $0$ through $6$ minus mean activity in bins $-6$ through $-1$. Two-sided randomization p-values use the standard plus-one adjustment. \autoref{fig_placebo_intensive_margin} reports the results.

\section{Price-Impact Estimation Details}
\label{appendix_impact}

\noindent The price-impact measures use the initiating order identified by the ledger. For initiating order $n$, let $x_n$ be its USDC value and let $s_n$ equal $+1$ when the order buys Trump YES or sells Trump NO and $-1$ when it sells Trump YES or buys Trump NO. A Trump YES transaction price enters directly, while a Trump NO transaction price $p_n^{NO}$ is mapped to the implied YES price $1-p_n^{NO}$. Within each 5-minute bin, we calculate the USDC-weighted average implied YES price and net signed flow in millions of USDC:
\begin{equation}
	p_t=\frac{\sum_{n\in t}x_np_n}{\sum_{n\in t}x_n},
	\qquad
	Q_t=\frac{1}{10^6}\sum_{n\in t}s_nx_n.
\end{equation}
If a bin contains no transaction, we carry forward the previous price and set $Q_t=0$. We transform the bounded price into log odds, $\theta_t=\log[p_t/(1-p_t)]$, and estimate
\begin{equation}
	\Delta\theta_t=\alpha+\lambda_t Q_t+\varepsilon_t
\end{equation}
and
\begin{equation}
	\Delta\theta_t=\alpha+\lambda_{\mathrm{perm},t}Q_t+\lambda_{\mathrm{trans},t}\Delta Q_t+\varepsilon_t.
\end{equation}
At each 5-minute point, both regressions use the trailing 2,016 bins, equivalent to seven calendar days. The Kyle confidence interval uses the OLS covariance estimate. The Glosten--Harris specification uses HAC standard errors with 12 lags. The displayed $[-2,+3]$ hour window traces how the coefficients change as event trades enter the trailing sample. These rolling estimates describe how executed order flow maps into prices around each event rather than identifying a causal effect of the shock on market depth.

\section{The Participation Specification}
\label{appendix_robustness}

\noindent Trading frequency can count several fills generated by one order. \autoref{tab_panel_intensity_5min} therefore uses a participation indicator equal to one when a trader is active in a 5-minute bin. The matched pre-event measure, \emph{Trad Int Multi}, indicates whether the trader was active in multiple pre-event bins. The main patterns are unchanged.

\section{Event-Specific Panel Regressions}
\label{appendix_event_specific_panel}

\noindent We estimate the full panel specification separately for each shock. \autoref{tab_event_volume_mtm}, \autoref{tab_event_freq_mtm}, and \autoref{tab_event_part_mtm} report the volume, frequency, and participation estimates. Prior volume is positive at all three events for every outcome. Realized gains are positively associated with the response at the assassination attempt and withdrawal, while the debate instead has positive loss-side coefficients. No single event therefore drives the pooled estimates in the main text.

\clearpage

\setcounter{section}{1}
\setcounter{table}{0}

\begin{table}[!htbp]
	\centering
	\caption{Representative ledger matches}
	\label{tab_fill_routes}
	\begin{minipage}{\textwidth}
	\centering
	\footnotesize
	\resizebox{\textwidth}{!}{%
	\begin{tabular}{rlllllrr}
		\toprule
		logIndex & event & wallet field & taker & gives & receives & amount given & amount received \\
		\midrule
		\multicolumn{8}{l}{\textit{Panel A: One resting order and one incoming order}} \\
		161 & OrderFilled & 0x9d84..1344 & 0xd42F..047d & USDC & Trump NO & 16.5731 & 28.09 \\
		163 & OrderFilled & 0xd42F..047d & \textbf{EXCHANGE} & Trump NO & USDC & 28.09 & 16.5731 \\
		164 & OrdersMatched & 0xd42F..047d & & Trump NO & USDC & 28.09 & 16.5731 \\
		\midrule
		\multicolumn{8}{l}{\textit{Panel B: One incoming order filled through direct exchange and minting}} \\
		300 & OrderFilled & 0x8698..35c1 & 0xf0b0..68ec & Trump YES & USDC & 200 & 84 \\
		314 & OrderFilled & 0xd42F..047d & 0xf0b0..68ec & USDC & Trump NO & 22.095238 & 38.095237 \\
		316 & OrderFilled & 0xf0b0..68ec & \textbf{EXCHANGE} & USDC & Trump YES & 99.999999 & 238.095237 \\
		317 & OrdersMatched & 0xf0b0..68ec & & USDC & Trump YES & 99.999999 & 238.095237 \\
		\bottomrule
	\end{tabular}
	}
	\par\vspace{2pt}
	\footnotesize\raggedright \textit{Notes:} Panel A reports block 51953304, transaction 0xa781a2...640e6. Panel B reports block 51951654, transaction 0x7fea7c...8c55. Wallets are truncated. ``Wallet field'' is \texttt{maker} for \texttt{OrderFilled} and \texttt{takerOrderMaker} for \texttt{OrdersMatched}. The \texttt{OrderFilled} row with \texttt{taker}=\textbf{EXCHANGE} identifies the incoming order. Amounts follow the units in the asset columns.
	\end{minipage}
\end{table}

\begin{table}[!htbp]
	\centering
	\caption{Net signed flow under ledger-initiator and tick-rule signing}
	\label{tab_dual_signing}
	\begin{threeparttable}
	\footnotesize
	\begin{tabular}{llrr}
		\toprule
		Event & Window & Ledger initiator & Tick rule \\
		\midrule
		Debate & $[0,+30\text{m})$ / $[0,+3\text{h})$ & $+11{,}234$ / $+54{,}580$ & $+9{,}439$ / $+45{,}587$ \\
		Assassination & $[0,+30\text{m})$ / $[0,+3\text{h})$ & $-205{,}953$ / $+156{,}486$ & $-233{,}033$ / $+91{,}484$ \\
		Withdrawal & $[0,+30\text{m})$ / $[0,+3\text{h})$ & $-208{,}389$ / $-202{,}599$ & $-203{,}618$ / $-168{,}789$ \\
		\bottomrule
	\end{tabular}
	\begin{tablenotes}[para,flushleft]
		\footnotesize
		\item \textit{Notes:} Net signed Trump YES flow in USDC, positive for exposure-raising pressure. Ledger signing uses the identified taker order. The tick rule signs the same Trump YES transactions from price changes.
	\end{tablenotes}
	\end{threeparttable}
\end{table}

\clearpage
\setcounter{section}{2}
\setcounter{figure}{0}

\begin{figure}[p]
	\centering
	\begin{subfigure}[t]{\textwidth}
		\centering
		\includegraphics[width=.32\linewidth]{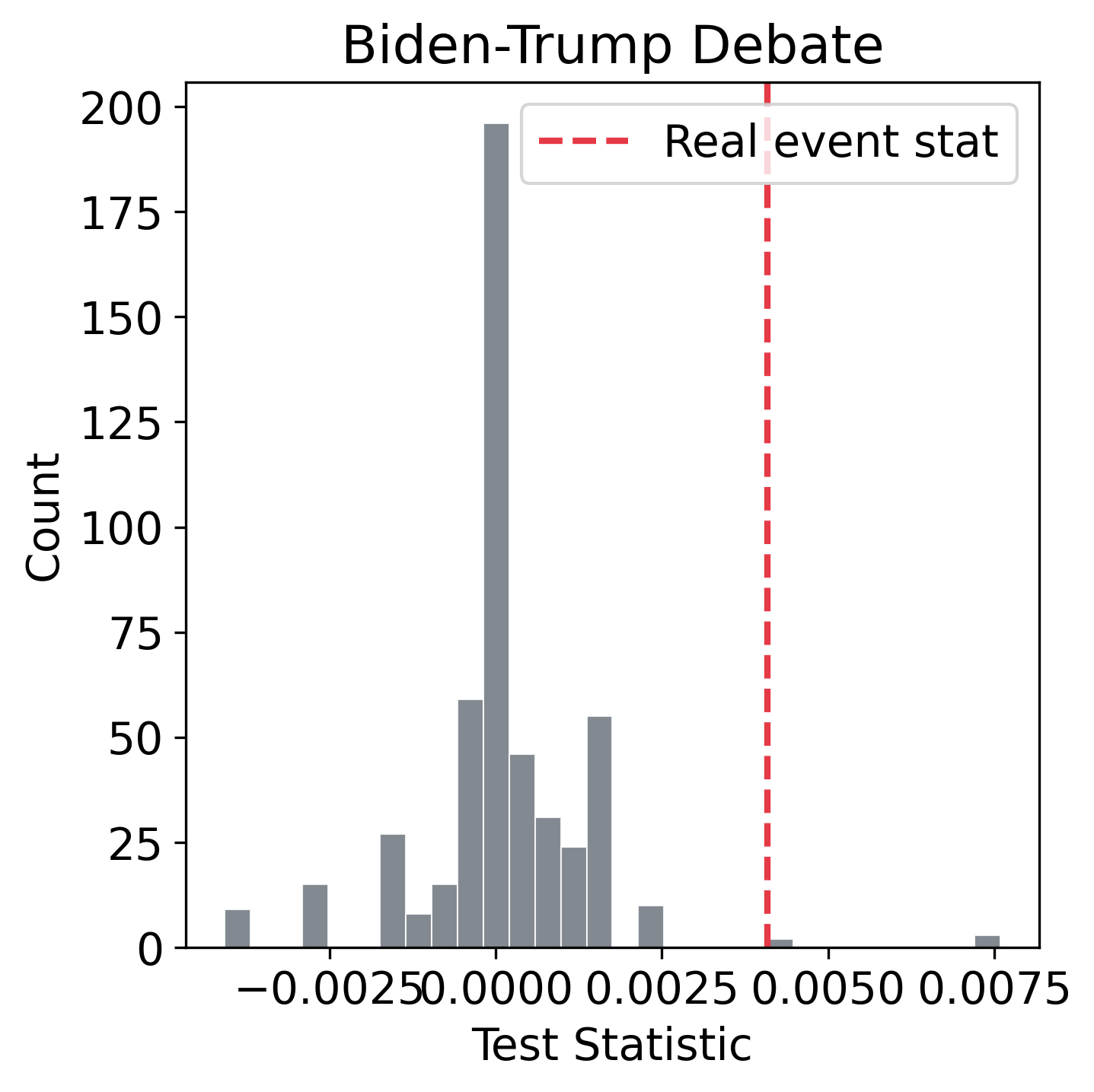}\hfill
		\includegraphics[width=.32\linewidth]{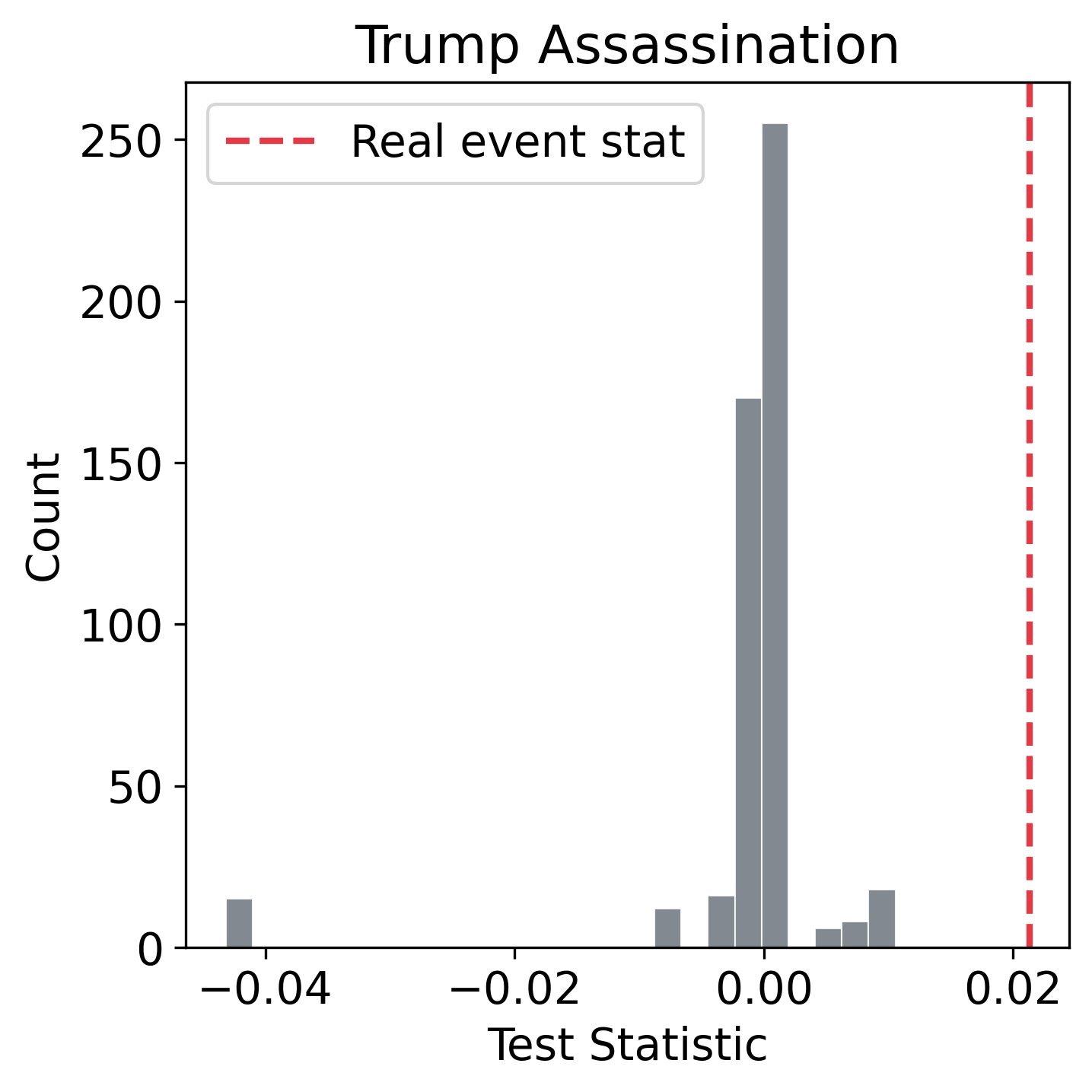}\hfill
		\includegraphics[width=.32\linewidth]{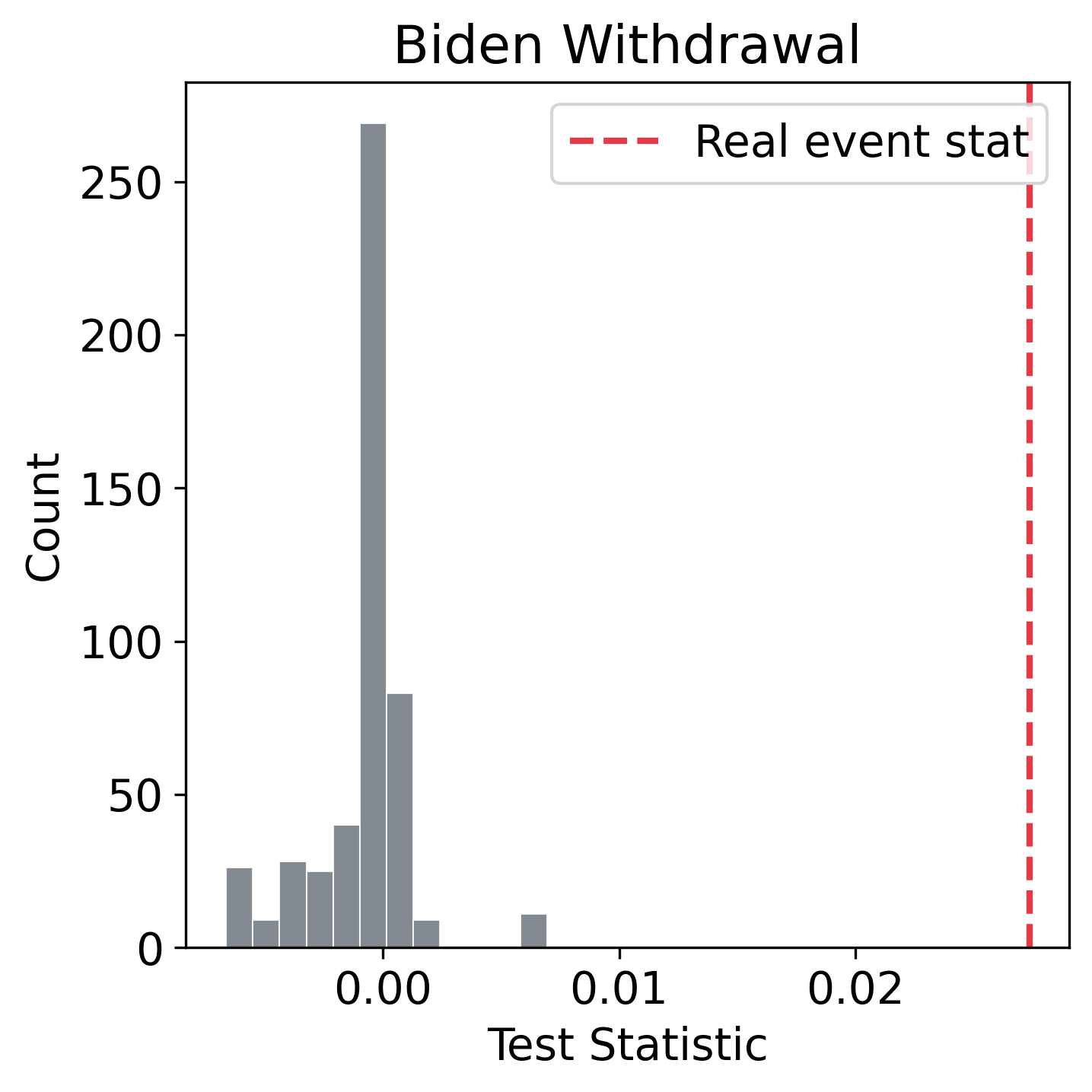}
		\caption{Abnormal incumbent trading volume}
	\end{subfigure}

	\vspace{0.5em}

	\begin{subfigure}[t]{\textwidth}
		\centering
		\includegraphics[width=.32\linewidth]{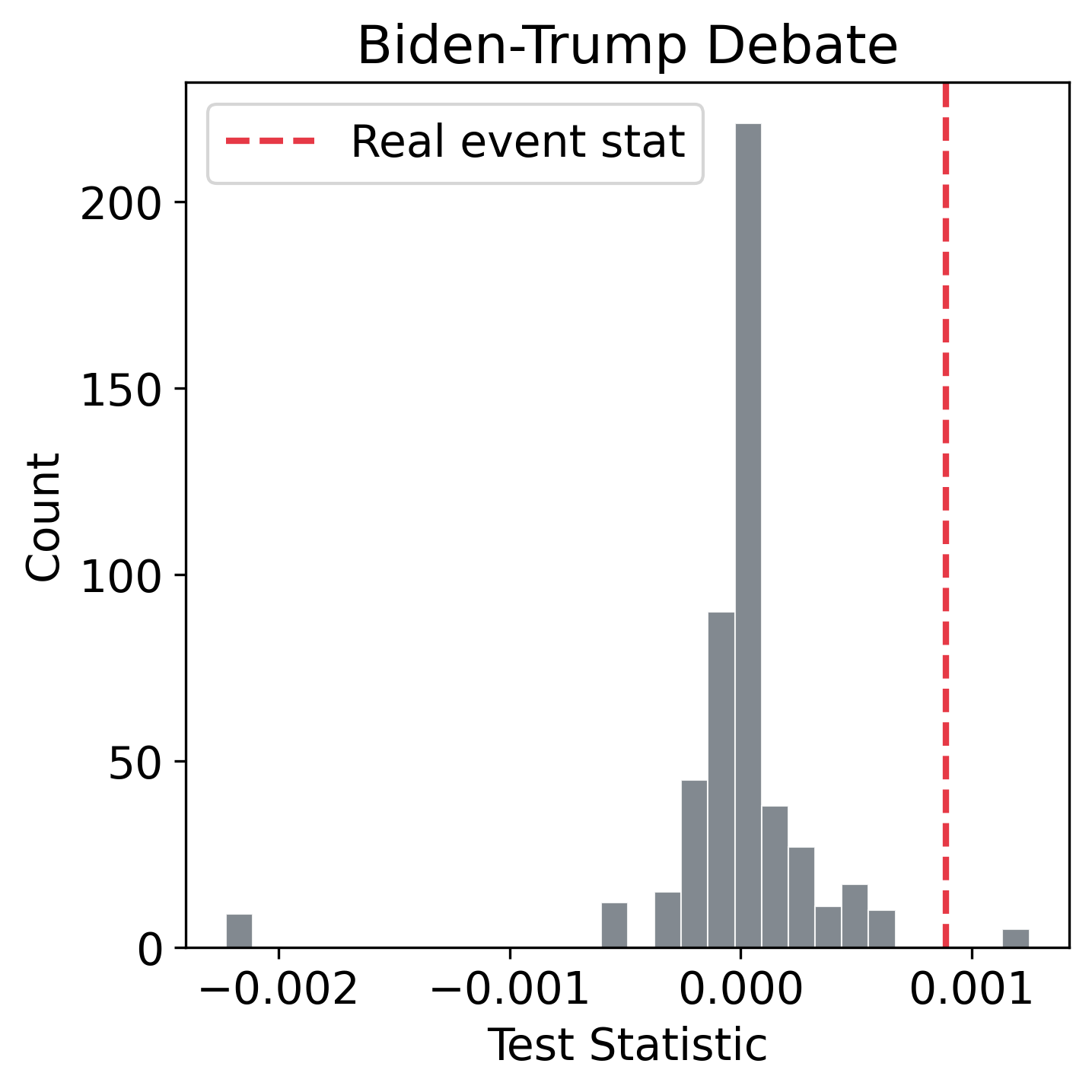}\hfill
		\includegraphics[width=.32\linewidth]{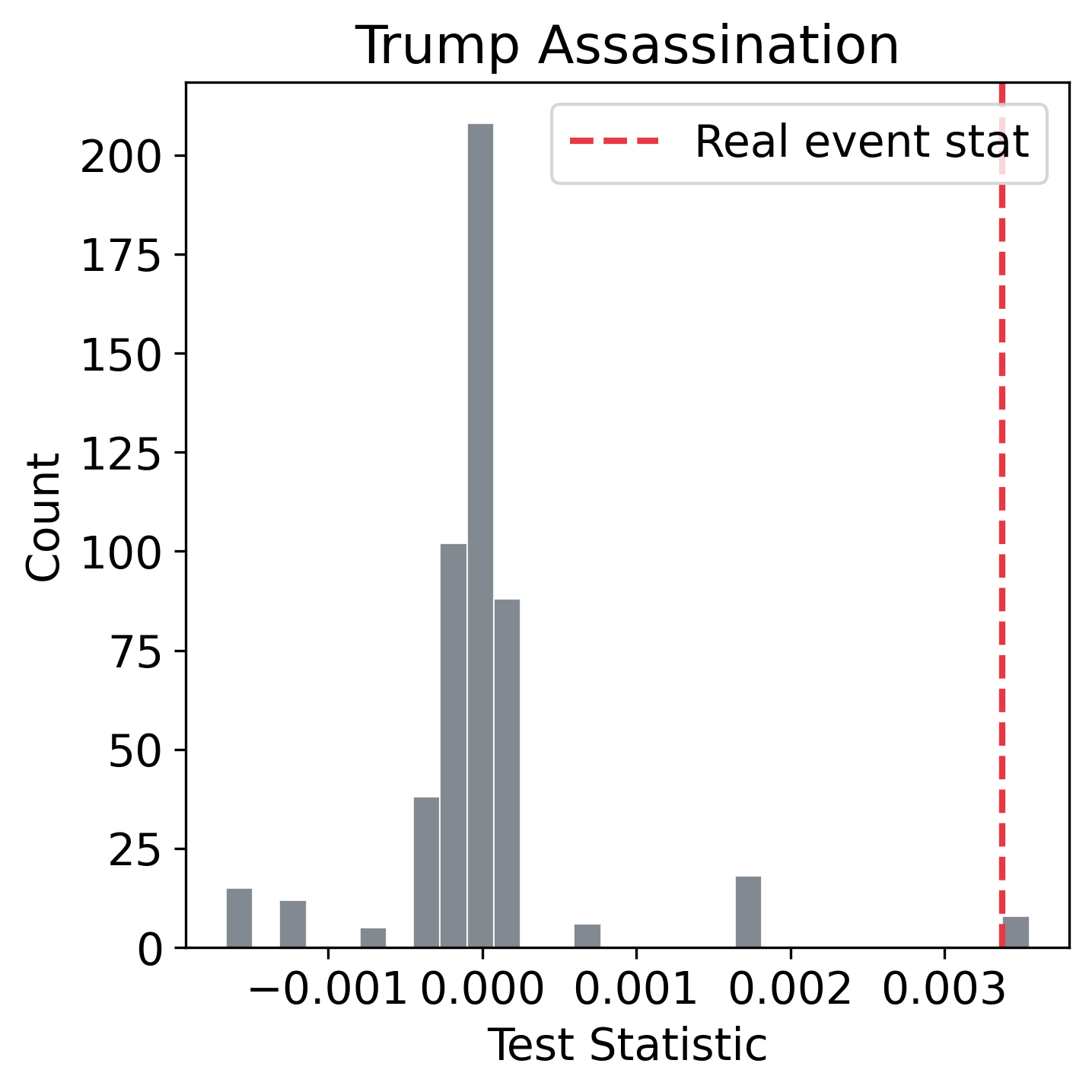}\hfill
		\includegraphics[width=.32\linewidth]{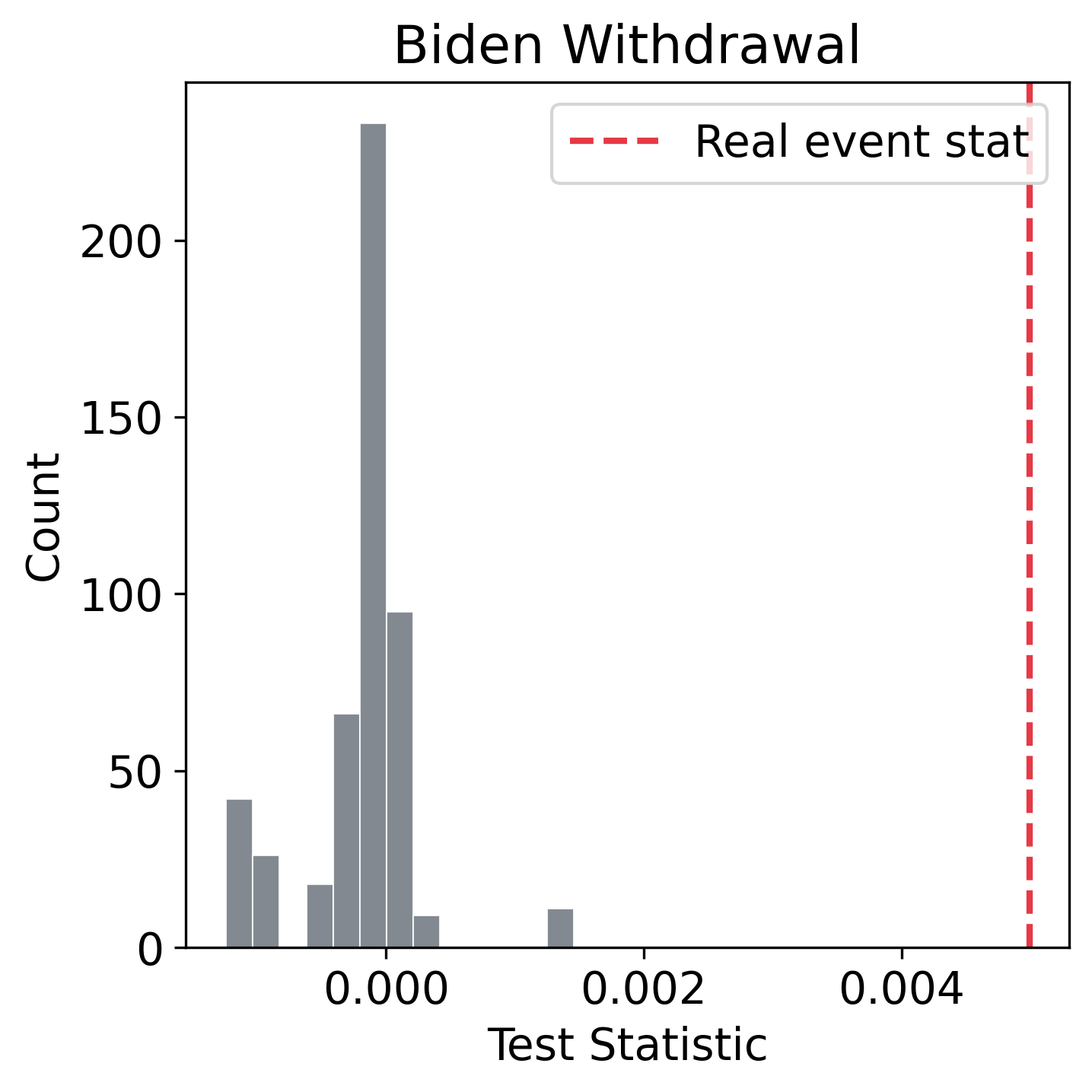}
		\caption{Abnormal incumbent trading frequency}
	\end{subfigure}

	\vspace{0.5em}

	\begin{subfigure}[t]{\textwidth}
		\centering
		\includegraphics[width=.32\linewidth]{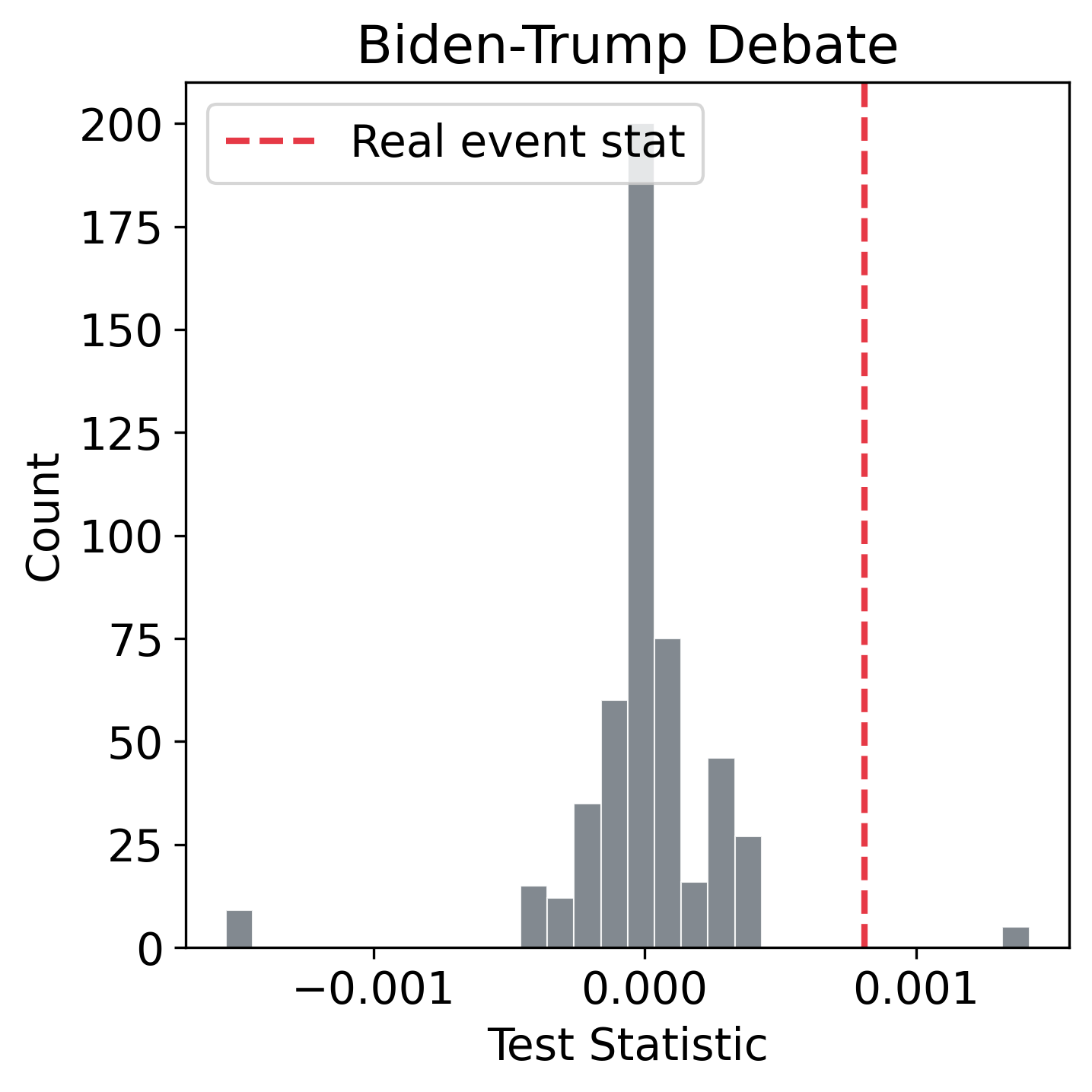}\hfill
		\includegraphics[width=.32\linewidth]{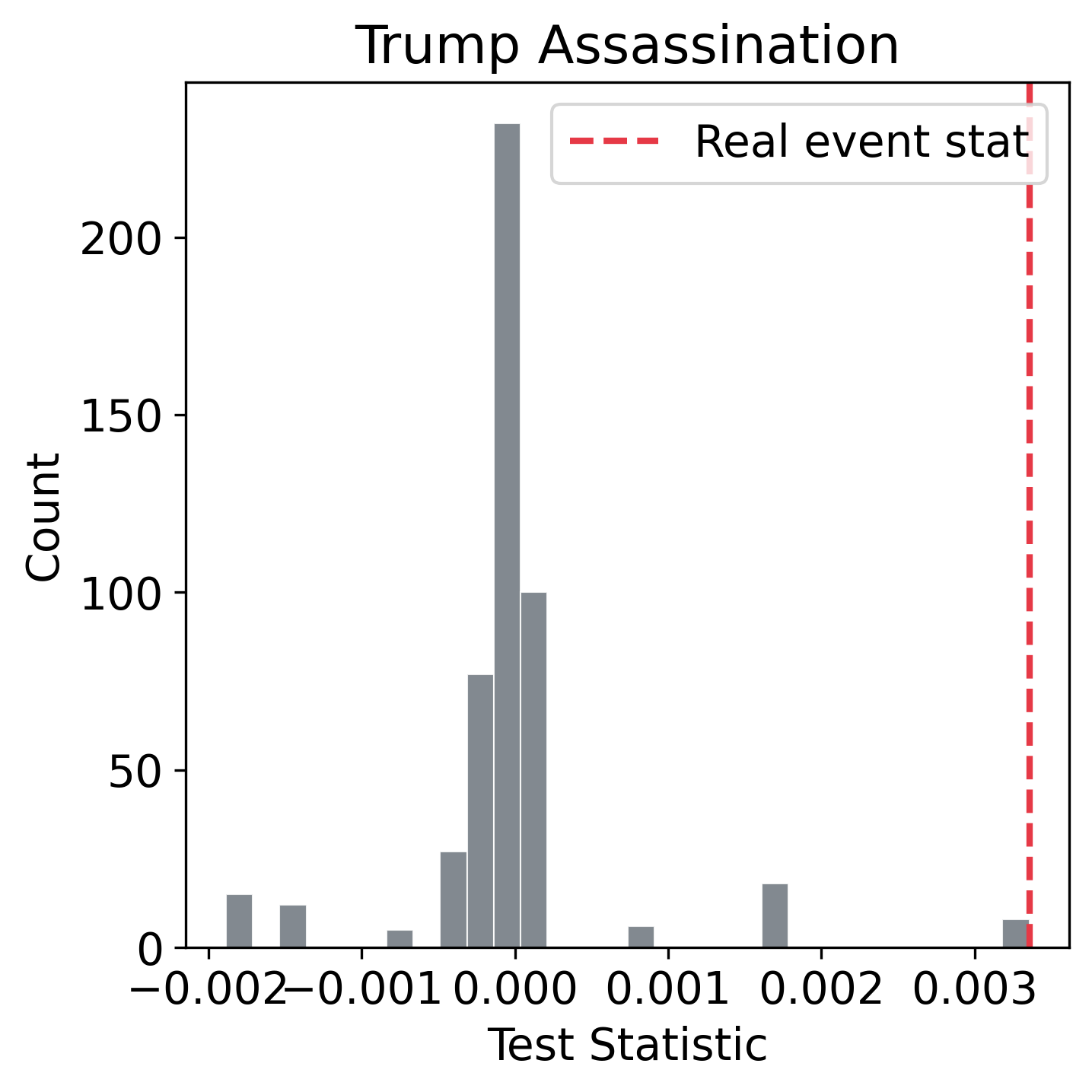}\hfill
		\includegraphics[width=.32\linewidth]{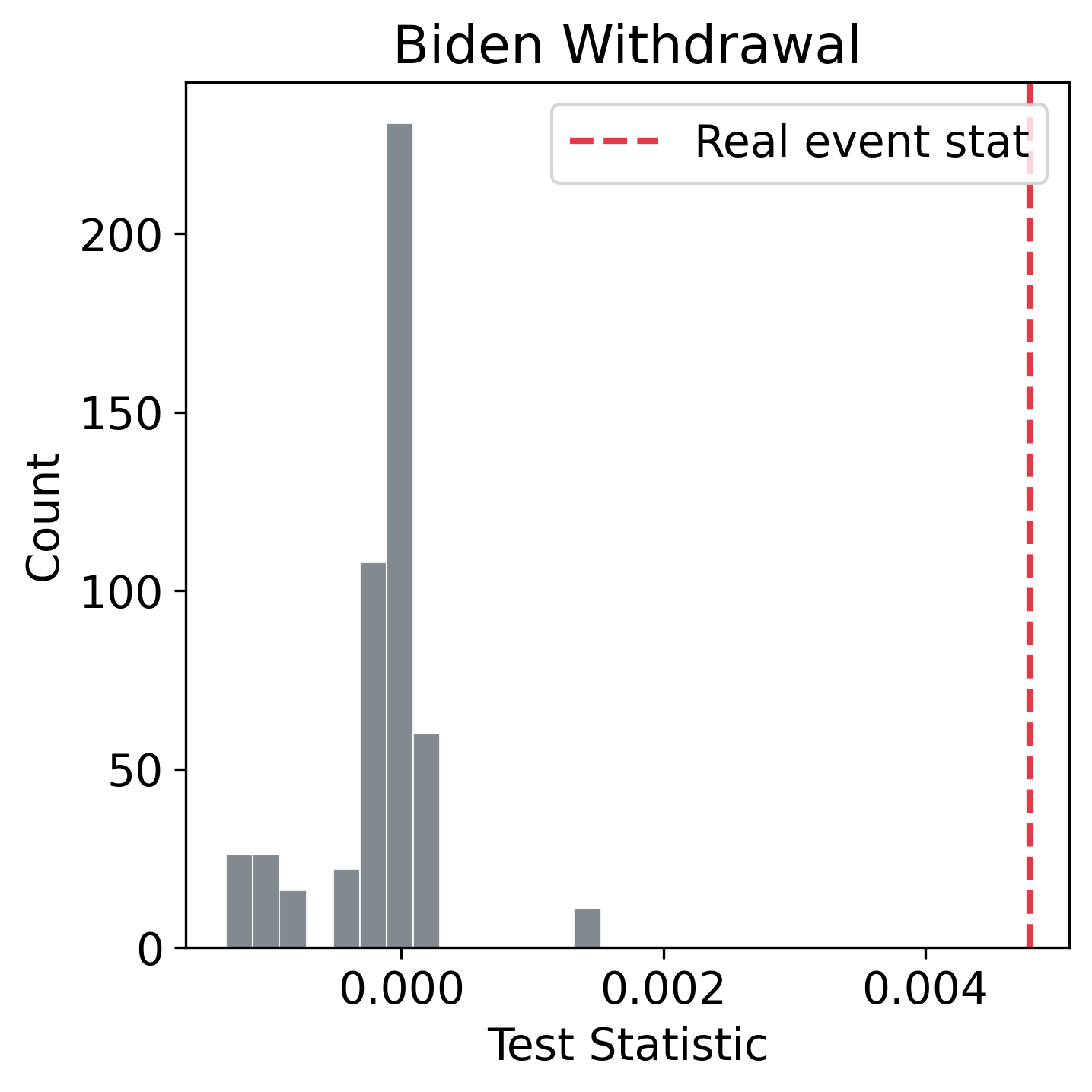}
		\caption{Abnormal incumbent trading participation}
	\end{subfigure}

	\caption{Placebo distributions for abnormal incumbent trading activity}
	\label{fig_placebo_intensive_margin}
	\vspace{0.25em}
	\begin{minipage}{0.98\linewidth}
	\footnotesize \raggedright \textit{Notes:} Histograms show 500 weekday-and-clock-time-matched placebo jumps. Dashed lines mark real-event values. Two-sided randomization $p$-values for debate/assassination/withdrawal are 0.012/0.032/0.002 for volume, 0.030/0.018/0.002 for frequency, and 0.030/0.002/0.002 for participation.
	\end{minipage}
\end{figure}

\clearpage

\setcounter{section}{4}
\setcounter{table}{0}

\begin{landscape}
\begin{table}[p]
\centering
\caption{Stacked panel regressions: trading participation}
\label{tab_panel_intensity_5min}
\begin{threeparttable}
\small
\setlength{\tabcolsep}{5pt}
\begin{tabular}{lccccccc}
	\toprule
	& (1) & (2) & (3) & (4) & (5) & (6) & (7) \\
	& Baseline & Trading Vol & Trad Int Multi & MtM Exposure & Single Market & Trader Types & Full \\
	\midrule
	Post Event & 0.0039*** & & & & & & \\
	& (0.0002) & & & & & & \\
	Post Event $\times$ Trad Vol High & & 0.0110*** & & & & & 0.0050*** \\
	& & (0.0006) & & & & & (0.0005) \\
	Post Event $\times$ Trad Int Multi & & & 0.0145*** & & & & 0.0115*** \\
	& & & (0.0009) & & & & (0.0010) \\
	Post Event $\times$ MtM Loss & & & & 0.0045*** & & & -0.0001 \\
	& & & & (0.0005) & & & (0.0005) \\
	Post Event $\times$ MtM Gain & & & & 0.0094*** & & & 0.0059*** \\
	& & & & (0.0007) & & & (0.0005) \\
		Post Event $\times$ Single Mkt & & & & & -0.0021** & & -0.0002 \\
	& & & & & (0.0010) & & (0.0010) \\
	Post Event $\times$ Contrarian & & & & & & 0.0049** & -0.0053** \\
	& & & & & & (0.0024) & (0.0025) \\
	Post Event $\times$ Momentum & & & & & & 0.0138*** & 0.0031 \\
	& & & & & & (0.0044) & (0.0045) \\
	\midrule
	Entity Effects & YES & YES & YES & YES & YES & YES & YES \\
	Time Effects & NO & YES & YES & YES & YES & YES & YES \\
	R-Squared (Within) & 0.0017 & 0.0046 & 0.0053 & 0.0034 & -0.0020 & 0.0006 & 0.0068 \\
		R-Squared (Between) & 0.0207 & 0.0497 & 0.0575 & 0.0356 & -0.0189 & 0.0059 & 0.0751 \\
		R-Squared (Overall) & 0.0040 & 0.0101 & 0.0116 & 0.0073 & -0.0041 & 0.0012 & 0.0151 \\
	Observations & 284420 & 284420 & 284420 & 284420 & 284420 & 284420 & 284420 \\
	\bottomrule
\end{tabular}
\begin{tablenotes}[para,flushleft]
\item
\footnotesize \raggedright \textit{Notes:} The dependent variable is a trader-level participation indicator equal to one if the trader trades in a 5-minute bin. The specification follows \autoref{tab_panel_volume_5min} and \autoref{tab_panel_freq_5min}. Standard errors, clustered by trader-event, are in parentheses. * p$<$0.1, ** p$<$0.05, *** p$<$0.01.
\end{tablenotes}
\end{threeparttable}
\end{table}
\end{landscape}

\setcounter{section}{5}
\setcounter{table}{0}

\begin{table}[p]
\centering
\caption{Event-specific panel regressions: trading volume}
\label{tab_event_volume_mtm}
\begin{threeparttable}
\small
\begin{tabular}{lcccc}
	\toprule
	& (1) & (2) & (3) & (4) \\
	& Debate & Assassination & Withdrawal & Pooled \\
	\midrule
	Post Event $\times$ Trad Vol High & 0.0063* & 0.0619*** & 0.0436*** & 0.0373*** \\
	& (0.0037) & (0.0078) & (0.0057) & (0.0035) \\
	Post Event $\times$ Trad Freq High & -0.0044 & 0.0275*** & 0.0583*** & 0.0480*** \\
	& (0.0042) & (0.0100) & (0.0069) & (0.0050) \\
	Post Event $\times$ MtM Loss & 0.0218*** & -0.0149*** & -0.0096*** & -0.0034 \\
	& (0.0050) & (0.0052) & (0.0036) & (0.0026) \\
	Post Event $\times$ MtM Gain & -0.0024 & 0.0246*** & 0.1053*** & 0.0417*** \\
	& (0.0030) & (0.0039) & (0.0101) & (0.0034) \\
	Post Event $\times$ Single Mkt & -0.0087 & 0.0004 & 0.0214** & 0.0047 \\
	& (0.0077) & (0.0098) & (0.0102) & (0.0058) \\
		Post Event $\times$ Contrarian & 0.0089 & 0.0989** & -0.0463 & -0.0147 \\
	& (0.0170) & (0.0407) & (0.0286) & (0.0160) \\
	Post Event $\times$ Momentum & 0.0433** & 0.0652 & 0.1179 & 0.0273 \\
	& (0.0177) & (0.0524) & (0.0859) & (0.0245) \\
	\midrule
	Entity Effects & YES & YES & YES & YES \\
	Time Effects & YES & YES & YES & YES \\
	R-Squared (Within) & 0.0016 & 0.0067 & 0.0095 & 0.0059 \\
	R-Squared (Between) & 0.0138 & 0.0703 & 0.1015 & 0.0657 \\
	R-Squared (Overall) & 0.0030 & 0.0136 & 0.0210 & 0.0130 \\
	Observations & 71581 & 93200 & 119639 & 284420 \\
	\bottomrule
\end{tabular}
\begin{tablenotes}[para,flushleft]
\item
\footnotesize \raggedright \textit{Notes:} Dependent variable: trader-level $\operatorname{asinh}(\text{trading volume})$ in 5-minute event time. Columns (1)--(3) are event-specific models with trader fixed effects. Column (4) reproduces the pooled model in \autoref{tab_panel_volume_5min} with trader-by-event fixed effects. All include event-time-bin fixed effects. Standard errors cluster by trader in (1)--(3) and trader-event in (4). * p$<$0.1, ** p$<$0.05, *** p$<$0.01.
\end{tablenotes}
\end{threeparttable}
\end{table}

\begin{table}[p]
\centering
\caption{Event-specific panel regressions: trading frequency}
\label{tab_event_freq_mtm}
\begin{threeparttable}
\small
\begin{tabular}{lcccc}
	\toprule
	& (1) & (2) & (3) & (4) \\
	& Debate & Assassination & Withdrawal & Pooled \\
	\midrule
	Post Event $\times$ Trad Vol High & 0.0013 & 0.0087*** & 0.0069*** & 0.0057*** \\
	& (0.0009) & (0.0012) & (0.0013) & (0.0007) \\
	Post Event $\times$ Trad Freq High & 0.0001 & 0.0061*** & 0.0112*** & 0.0092*** \\
	& (0.0009) & (0.0017) & (0.0015) & (0.0010) \\
	Post Event $\times$ MtM Loss & 0.0039*** & -0.0023** & -0.0000 & 0.0004 \\
	& (0.0011) & (0.0010) & (0.0008) & (0.0006) \\
	Post Event $\times$ MtM Gain & -0.0004 & 0.0034*** & 0.0165*** & 0.0064*** \\
	& (0.0007) & (0.0007) & (0.0020) & (0.0006) \\
	Post Event $\times$ Single Mkt & -0.0019 & 0.0001 & 0.0011 & -0.0003 \\
	& (0.0022) & (0.0018) & (0.0024) & (0.0013) \\
	Post Event $\times$ Contrarian & 0.0020 & 0.0120** & -0.0080* & -0.0035 \\
	& (0.0037) & (0.0059) & (0.0043) & (0.0028) \\
	Post Event $\times$ Momentum & 0.0083** & 0.0120 & 0.0234 & 0.0061 \\
	& (0.0040) & (0.0107) & (0.0184) & (0.0052) \\
	\midrule
	Entity Effects & YES & YES & YES & YES \\
	Time Effects & YES & YES & YES & YES \\
	R-Squared (Within) & 0.0014 & 0.0060 & 0.0095 & 0.0058 \\
	R-Squared (Between) & 0.0126 & 0.0624 & 0.0844 & 0.0562 \\
	R-Squared (Overall) & 0.0028 & 0.0123 & 0.0203 & 0.0125 \\
	Observations & 71581 & 93200 & 119639 & 284420 \\
	\bottomrule
\end{tabular}
\begin{tablenotes}[para,flushleft]
\item
\footnotesize \raggedright \textit{Notes:} Dependent variable: trader-level $\operatorname{asinh}(\text{trading frequency})$ in 5-minute event time. Columns (1)--(3) are event-specific models with trader fixed effects. Column (4) reproduces the pooled model in \autoref{tab_panel_freq_5min} with trader-by-event fixed effects. All include event-time-bin fixed effects. Standard errors cluster by trader in (1)--(3) and trader-event in (4). * p$<$0.1, ** p$<$0.05, *** p$<$0.01.
\end{tablenotes}
\end{threeparttable}
\end{table}

\begin{table}[p]
\centering
\caption{Event-specific panel regressions: trading participation}
\label{tab_event_part_mtm}
\begin{threeparttable}
\small
\begin{tabular}{lcccc}
	\toprule
	& (1) & (2) & (3) & (4) \\
	& Debate & Assassination & Withdrawal & Pooled \\
	\midrule
	Post Event $\times$ Trad Vol High & 0.0009 & 0.0080*** & 0.0057*** & 0.0050*** \\
	& (0.0006) & (0.0011) & (0.0009) & (0.0005) \\
		Post Event $\times$ Trad Int Multi & 0.0001 & 0.0067*** & 0.0153*** & 0.0115*** \\
	& (0.0009) & (0.0016) & (0.0014) & (0.0010) \\
	Post Event $\times$ MtM Loss & 0.0040*** & -0.0027*** & -0.0007 & -0.0001 \\
	& (0.0009) & (0.0009) & (0.0007) & (0.0005) \\
	Post Event $\times$ MtM Gain & -0.0002 & 0.0032*** & 0.0148*** & 0.0059*** \\
	& (0.0005) & (0.0006) & (0.0016) & (0.0005) \\
	Post Event $\times$ Single Mkt & -0.0007 & -0.0000 & 0.0017 & -0.0002 \\
	& (0.0015) & (0.0015) & (0.0018) & (0.0010) \\
	Post Event $\times$ Contrarian & 0.0023 & 0.0121** & -0.0117*** & -0.0053** \\
	& (0.0033) & (0.0058) & (0.0029) & (0.0025) \\
	Post Event $\times$ Momentum & 0.0075** & 0.0117 & 0.0176 & 0.0031 \\
	& (0.0032) & (0.0102) & (0.0152) & (0.0045) \\
	\midrule
	Entity Effects & YES & YES & YES & YES \\
	Time Effects & YES & YES & YES & YES \\
	R-Squared (Within) & 0.0019 & 0.0067 & 0.0112 & 0.0068 \\
	R-Squared (Between) & 0.0200 & 0.0753 & 0.1158 & 0.0751 \\
	R-Squared (Overall) & 0.0041 & 0.0141 & 0.0246 & 0.0151 \\
	Observations & 71581 & 93200 & 119639 & 284420 \\
	\bottomrule
\end{tabular}
\begin{tablenotes}[para,flushleft]
\item
\footnotesize \raggedright \textit{Notes:} Dependent variable: indicator for trader activity in a 5-minute bin. Columns (1)--(3) are event-specific models with trader fixed effects. Column (4) reproduces the pooled model in \autoref{tab_panel_intensity_5min} with trader-by-event fixed effects. All include event-time-bin fixed effects. Standard errors cluster by trader in (1)--(3) and trader-event in (4). * p$<$0.1, ** p$<$0.05, *** p$<$0.01.
\end{tablenotes}
\end{threeparttable}
\end{table}

\end{document}